\documentclass[english,prl, twocolumn, longbibliography, superscriptaddress]{revtex4-1}
\usepackage[T1]{fontenc}
\usepackage{array}
\usepackage{verbatim}
\usepackage{float}
\usepackage{amsmath}
\usepackage{graphicx}

\makeatletter

\providecommand{\tabularnewline}{\\}

\usepackage{amsmath}
\usepackage{subfigure}
\usepackage{graphicx, mathtools}
\usepackage[colorlinks=true,linkcolor=MidnightBlue,urlcolor=MidnightBlue,citecolor=MidnightBlue,anchorcolor=MidnightBlue]{hyperref}\usepackage[dvipsnames]{xcolor}

\makeatother

\usepackage{babel}
\begin{document}
\title{Periodic orbits of active particles induced by hydrodynamic monopoles}
\author{Austen Bolitho}
\email{ab2075@cam.ac.uk }

\affiliation{DAMTP, Centre for Mathematical Sciences, University of Cambridge,
Cambridge CB3 0WA, United Kingdom}
\author{Rajesh Singh}
\email{rs2004@cam.ac.uk}

\affiliation{DAMTP, Centre for Mathematical Sciences, University of Cambridge,
Cambridge CB3 0WA, United Kingdom}
\author{R. Adhikari}
\email{ra413@cam.ac.uk, rjoy@imsc.res.in}

\affiliation{DAMTP, Centre for Mathematical Sciences, University of Cambridge,
Cambridge CB3 0WA, United Kingdom}
\affiliation{The Institute of Mathematical Sciences-HBNI, CIT Campus, Chennai 600113,
India}
\begin{abstract}
Terrestrial experiments on active particles, such as \emph{Volvox},
involve gravitational forces, torques and accompanying monopolar fluid
flows. Taking these into account, we analyse the dynamics of a pair
of self-propelling, self-spinning active particles between widely separated
parallel planes. Neglecting flow reflected by the planes, the dynamics
of orientation and horizontal separation is symplectic, with a Hamiltonian
exactly determining limit cycle oscillations. Near the bottom plane,
gravitational torque damps and reflected flow excites this oscillator,
sustaining a second limit cycle that can be perturbatively related
to the first. Our work provides a theory for dancing \emph{Volvox}
and highlights the importance of monopolar flow in active matter.
\end{abstract}
\maketitle
Since Lighthill's seminal work on the squirming motion of a sphere
\citep{lighthill1952,blake1971a}, it has been understood that freely
moving active particles produce hydrodynamic flows that disallow monopoles
and antisymmetric dipoles \citep{anderson1989colloid}. The minimal representation
of active flows by the symmetric dipole, the leading term consistent
with force-free, torque free-motion, has been the basis of much theoretical
work in both particle \citep{lauga2009,drescher2010,koch2011} and
field representations of active matter \citep{marchetti2013,ramaswamy2017active}.
The importance of multipoles beyond leading order in representing
experimentally measured flows around active particles has now been
recognized and their effects have been included in recent theoretical
work \citep{Ghose2014,Pedley2016}. Less recognised, however, is the
fact that active particles in typical experiments \citep{drescher2009,drescher2010,goldstein2015green,palacci2013living,palacci2010sedimentation,buttinoni2013DynamicClustering,ebbens2010pursuit}
are neither force- nor torque-free: mismatches between particle and
solvent densities lead to net gravitational forces while mismatches
between the gravitational and geometric centers lead to net gravitational
torques. In this case, both monopolar and antisymmetric dipolar flows are
allowed and become dominant, at long distances, over active contributions.
It is of great interest, therefore, to understand how these components
influence the dynamics of active particles and, more generally, of
active matter.

Theoretical work on this aspect of active matter has been limited, even
though the effect of monopolar flow in passive, driven matter is well-understood \citep{smoluchowski1911mutual, brenner1963, ladd1988,squiresWall, squires2001effective}.
Attraction induced by monopolar flow  near boundaries
has been shown to cause crystallisation of active particles \citep{singh2016crystallization}.
Reorientation induced  by monopolar vorticity has been identified as the
key mechanism in the emergence  of the pumping state of harmonically confined active particles \citep{nash2010,singh2015many}. However, none
of these studies have focused on the dynamics of pairs, which forms
the foundation for understanding collective motion, or attempted an
analytical description of motion.

In this Letter, we provide a theory for the dynamics of density-mismatched,
bottom-heavy, self-propelling and self-spinning active particles between widely separated
	parallel planes. Starting from the ten-dimensional equations
for hydrodynamically interacting active motion in the presence of external
forces and torques, we derive, by exploiting symmetries, a lower-dimensional
dynamical system for the pair. For positive buoyant mass, negative
gravitaxis, and negligible reflected flow, we obtain a sedimenting
state with limit cycle oscillations in the relative orientation and
horizontal separation. The dynamics is symplectic and a Hamiltonian
completely determines the properties of periodic orbits. On approach
to the bottom wall, reflected flow arrests sedimentation and yields
a levitating state with limit cycle oscillations that now includes
the mean height. This second limit cycle can be understood as a damped
(by gravitational torque) and driven (by reflected flow) perturbation
of the first. These rationalise the \emph{Volvox} dance \citep{drescher2009,goldstein2015green}
and highlight the importance of monopolar hydrodynamic flow in active
matter. We now explain how our results are obtained.
\begin{figure*}
\includegraphics[width=1\linewidth]{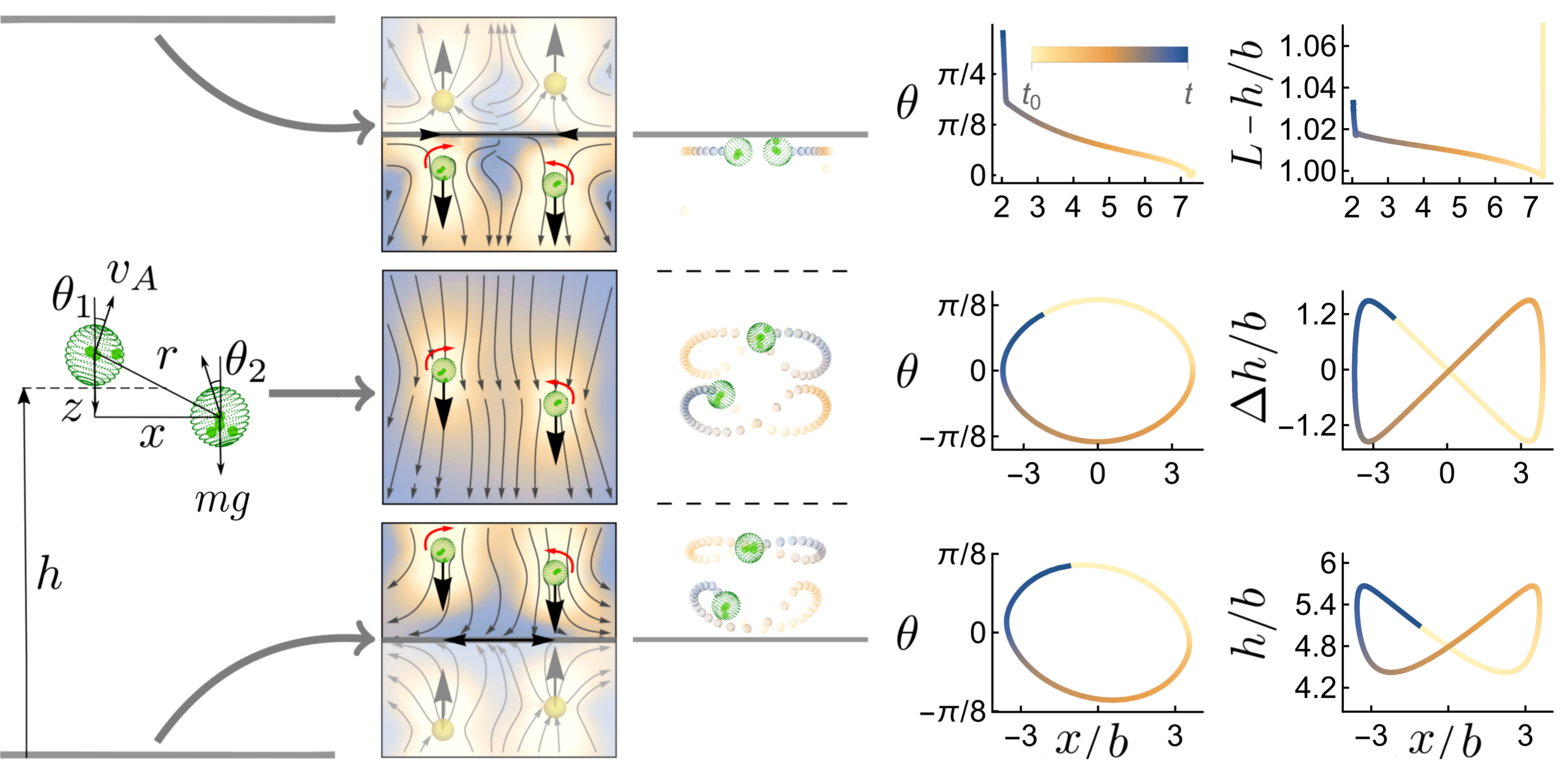}

\caption{Fixed point and limit cycles of the five-dimensional dynamics system
of Eq.(\ref{eq:dynamicalSystem-1}-\ref{eq:dynamicalSystem}). \emph{First
column}: coordinate system used to describe a pair of particles between
two parallel plane surfaces. \emph{Second column: }streamlines of
the monopolar flow (red curved arrows show flow-induced rotations
of the particles) superimposed on a pseudo colormap of the flow speed.
The plots correspond to the following three
cases: \emph{top row}: near the top surface, \emph{middle row}: away
from the surfaces, and \emph{bottom row}: near the bottom surface.
\emph{Third column}: stroboscopic images of the two-particle dynamics
in the three configurations. The dynamical system admits a fixed point
at the top surface, while limit cycles are formed away from the surfaces
and near the bottom surface. The last two columns contain the plot
of relative orientation $\theta$ and average height $h$ as a function
of $x$ for the three cases. The colorbar indicates time in the final 3 columns and $L$ is the separation of the planes. See movie 1 of SI. \label{fig:flagship-pic}}
\end{figure*}

\emph{Full and reduced equations}: We consider a pair of spherical
active particles of radius $b$, density $\rho$, self-propulsion
speed $v_{A}$, and self-rotation speed $\omega_{A}$, in an incompressible
Newtonian fluid of density $\rho_{f}$ and viscosity $\eta$
between parallel planes whose separation is $L\gg b$ \citep{greenFuncApprox}. Their geometric
centres, propulsive orientations, velocities, and angular velocities
are, respectively, $\boldsymbol{R}_{i}$, $\boldsymbol{p}_{i}$, $\boldsymbol{V}_{i}$
and $\boldsymbol{\Omega}_{i}$, where $i=1,2$ is the particle index.
Overdamped, hydrodynamically interacting, active motion in the presence
of body forces $\boldsymbol{F}_{j}^{B}$ and body torques $\boldsymbol{T}_{j}^{B}$
is given by \citep{singh2018generalized}\begin{equation}\label{eq:RBM}
\begin{aligned}
\boldsymbol{V}_{i}=\boldsymbol{\mu}_{ij}^{TT}\cdot\boldsymbol{F}_{j}^{B} & +\boldsymbol{\mu}_{ij}^{TR}\cdot\boldsymbol{T}_{j}^{B}+v_{A}\boldsymbol{p}_{i},\\
\boldsymbol{\Omega}_{i}=\boldsymbol{\mu}_{ij}^{RT}\cdot\boldsymbol{F}_{j}^{B} & +\boldsymbol{\mu}_{ij}^{RR}\cdot\boldsymbol{T}_{j}^{B}+\omega_{A}\boldsymbol{p}_{i}
\end{aligned}
\end{equation}where $\boldsymbol{\mu}_{ij}^{\alpha\beta}$ are
mobility matrices and repeated particle indices are summed. Positions
and orientations obey the kinematic equations $\dot{\boldsymbol{R}}_{i}=\mathbf{V}_{i},$
and $\dot{\boldsymbol{p}}_{i}=\mathbf{\boldsymbol{\Omega}}_{i}\times\boldsymbol{p}_{i}$.
The above follow directly from Newton's laws for active particles
when inertia and active flows are neglected \citep{siText}. The expression
for the exterior fluid flow $\boldsymbol{v}(\boldsymbol{r})$ around
the colloids is then: $\boldsymbol{v}(\boldsymbol{r})=(1+\tfrac{b^{2}}{6}\nabla^{2})\,\mathbf{G}\cdot\boldsymbol{F}_{i}^{B}+\frac{1}{2}(\boldsymbol{\nabla}\times\mathbf{G})\cdot\boldsymbol{T}_{i}^{B}$,
where $\mathbf{G}$ is a Green's function of Stokes equation \citep{pozrikidis1992}
which satisfies the appropriate boundary conditions at the boundaries
in the flow. The flow due to self-propulsion and self-spin involve, respectively,
two and three gradients of the Green's function and are thus subdominant \citep{Ghose2014,singh2018generalized}.
For a sphere in a gravitational field $\boldsymbol{g}$,
the force is $\boldsymbol{F}_{i}^{B}$=$m\boldsymbol{g}$, where $m$=$\frac{4\pi b^{3}}{3}$$(\rho-\rho_{f})$
is the buoyant mass and the torque is $\boldsymbol{T}_{i}^{B}$=$\boldsymbol{d}_{i}\times(\frac{4\pi b^{3}}{3}\rho\boldsymbol{g})$,
where $\boldsymbol{d}_{i}$ is the position of the centre of gravity
relative to $\boldsymbol{R}_{i}$ \citep{pedley1992}. The torque
aligns $\boldsymbol{d}_{i}$ parallel to $\boldsymbol{g}$ and positive/negative
gravitaxis results when $\boldsymbol{p}_{i}$ is parallel/anti-parallel
to $\boldsymbol{d}_{i}$. Typical estimates of these parameters for
a $Volvox$ are $b\sim300$$\mathrm{\mu m}$, $v_{A}\sim300$$\mathrm{\mu m/s}$,
$mg\sim1\mathrm{nN}$, $\omega_{A}\sim1$ rad/s \citep{drescher2009}.
Thus, the typical active forces $F^{A}\sim6\pi\eta bv_{A}\sim10^{-9}$N
and torques $T^{A}\sim8\pi\eta b^{3}\omega^{A}\sim10^{-12}$Nm. Thus,
Brownian forces $k_{B}T/b\sim10^{-14}$N and torques $k_{B}T\sim10^{-20}$Nm
can be neglected for such systems of active particles. We now present
a reduced description of our deterministic equations of motion.

Our dimensional reduction is motivated by a symmetry of Stokes flow
that constrains motion initially in a plane perpendicular to the torque
to remain in that plane. We choose $y=0$ to be the plane of motion,
set $\text{\ensuremath{\boldsymbol{F}_{i}^{B}=-mg\hat{\boldsymbol{z}}}}$,
$\boldsymbol{T}_{i}^{B}=T_{R}\boldsymbol{p}_{i}\times\hat{\boldsymbol{z}}$,
where $T_{R}=\frac{4\pi b^{3}}{3}\rho gd$ is the magnitude of the
gravitational torque, and parametrise $\boldsymbol{R}_{i}=x_{i}\hat{\boldsymbol{x}}+z_{i}\hat{\boldsymbol{z}}$
and $\boldsymbol{p}_{i}=\sin\theta_{i}\hat{\boldsymbol{x}}+\cos\theta_{i}\hat{\boldsymbol{z}}$,
so that $\boldsymbol{V}_{i}=\dot{x}_{i}\hat{\boldsymbol{x}}+\dot{z}_{i}\hat{\boldsymbol{z}},$
and $\boldsymbol{\Omega}_{i}=\dot{\theta}_{i}\hat{\boldsymbol{y}}$.
Using these and translational and time-reversal symmetries in Eq.(\ref{eq:RBM}),
retaining terms in the mobility matrices to leading order in $x_{1}-x_{2}$,
$z_{1}$ and $z_{2}$, discarding the decoupled equation for the horizontal
component of the center of mass, and expressing the result in terms
of the reduced variables $2\dot{\psi}=\dot{\theta}_{1}+\dot{\theta}_{2}$,
$2\dot{\theta}=\dot{\theta}_{1}-\dot{\theta}_{2}$, $x=x_{1}-x_{2},$
$z=z_{1}-z_{2}$, $2h=z_{1}+z_{2}$, we obtain a five-dimensional
dynamical system \citep{siText}, partitioned into two orientational
equations
\begin{subequations}\label{eq:dynamicalSystem-1}
\begin{alignat}{1}
\dot{\psi} & =\mkern-3mu-\frac{T_{R}}{8\pi\eta b^{3}}\sin\psi\cos\theta,\\
\dot{\theta} & =\mkern-3mu-\frac{T_{R}}{8\pi\eta b^{3}}\cos\psi\sin\theta-\frac{mg}{8\pi\eta}\left[\frac{x}{r^{3}}\mkern-3mu-\mkern-4mu\frac{x}{(4h^{2}\mkern-3mu+\mkern-3mur^{2})^{3/2}}\right]
\end{alignat}
\end{subequations}
and three positional equations,
\begin{subequations}\label{eq:dynamicalSystem}
\begin{alignat}{1}
\dot{x} & =\,\,\,\,2v_{A}\cos\psi\sin\theta+\frac{mghx}{2\pi\eta(4h^{2}+r^{2})^{3/2}},\\
\dot{z} & =-2v_{A}\sin\psi\cos\theta-\frac{mgz}{2\pi\eta(4h^{2}-z^{2})},
\\
\dot{h} & =v_{A}\cos\psi\cos\theta-\frac{mg}{8\pi\eta}\left(\frac{4}{3b}+\frac{1}{r}+\frac{z^{2}}{r^{3}}-\frac{2}{h}\right).
\end{alignat}
\end{subequations}
The geometry of the reduced variables is shown in Fig. (\ref{fig:flagship-pic}).
The orientational equations describe the competition between gravitational
torques that restore vertical orientations \citep{noteStability} and hydrodynamic torques,
from monopolar vorticity, that promotes relative re-orientation.
The first and second positional equations describe the change in relative
separation due to gravitaxis and reflected monopolar flow, the latter
of which increases horizontal separation and decreases vertical separation \citep{squires2001effective}.
The third positional equation describes the competition between the
tendency of the mean height to increase, due to gravitaxis and reflected
monopolar flow, and its tendency to decrease, due to gravitational
forces and monopolar flow. Eqs.(\ref{eq:dynamicalSystem-1}-\ref{eq:dynamicalSystem}) describe the sedimentation
of a pair of passive particles when $v_{A},\omega_{A}=0$ \citep{smoluchowski1911mutual};
the horizontal dynamics of a pair of phoretic particles when $v_{A}\neq0,\omega_{A}=0$
and both the height and orientation are fixed \citep{squires2001effective}; and the coupled dynamics
of horizontal separation and relative orientation when $v_{A}\neq0,\omega_{A}\ne0$
and the height is fixed \citep{drescher2009}.  

\begin{figure}
\includegraphics[width=1\columnwidth]{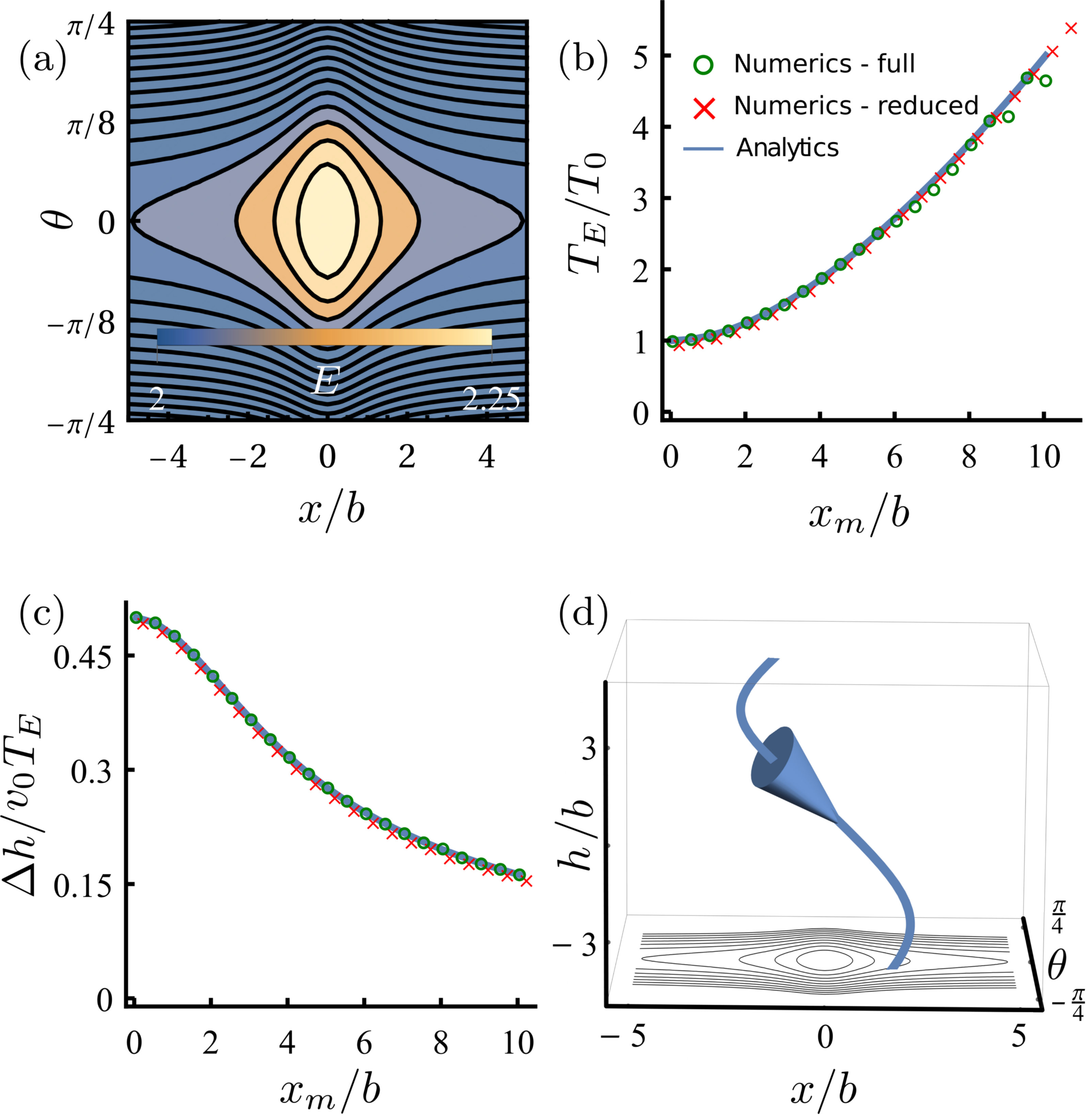}

\caption{\label{fig:unboundedPlots} Exact solution of the Hamiltonian limit
cycle. (a) level sets of the Hamiltonian in which the reflection
symmetries $x\rightarrow-x$ and $\theta\rightarrow-\theta$ are clearly
visible, (b) variation of the period of large oscillations $T_{E}$
as a function of their amplitude scaled by the period of small oscillations.
Using experimental values \citep{drescher2009}, we obtain $T_{0}\sim8$s
giving $T_{E}\sim12$s for $x_{m}=z=3$. This agrees well
with the experimentally measured time period of the \textquotedblleft minuet\textquotedblright{}
bound state \citep{drescher2009}, (c) sedimentation speed as a function
of the oscillation amplitude, and (d) orbit on a constant ``energy'' manifold
in $x,\theta$ and $h$ with the level sets of $H$ shown on a cross-section.
Exact analytical results are compared with numerical simulations of
the full (Eq.(\ref{eq:RBM})) and reduced (Eqs.(\ref{eq:dynamicalSystem-1}b,\ref{eq:dynamicalSystem}a,\ref{eq:dynamicalSystem}c))} equations in (b) and (c) with $T_R,\,1/h=0$.\vspace{-.6cm}
\end{figure}

\emph{Hamiltonian limit cycle}: We now analyse Eqs.(\ref{eq:dynamicalSystem-1}-\ref{eq:dynamicalSystem}), initially neglecting the reflected flow. We assume initial heights that are remote from both planes, $0\ll z_{1},z_{2}\ll L$
and parameter values, to be identified below, that ensure sedimentation
in the mean. The attractor $\psi=0$ of the first orientational equation,
reached on the time scale $\omega_{R}=T_{R}/8\pi\eta b^{3}$, defines
the slow manifold $\theta_{1}+\theta_{2}=0$. On this slow manifold
and neglecting reflected flow, reorientation is principally due to
the monopolar vorticity, $\dot{\theta}=-mgx/8\pi\eta r^{3}$, relative
horizontal motion is principally due to gravitaxis, $\dot{x}=2v_{A}\sin\theta$,
and relative vertical motion is absent, $\dot{z}=0.$ Remarkably,
the dynamics, which are governed by the reduced equations (\ref{eq:dynamicalSystem-1}b,\ref{eq:dynamicalSystem}a,\ref{eq:dynamicalSystem}c), has the symplectic form $\dot{x}=-\partial_{\theta}H$,
$\dot{\theta}=\partial_{x}H$ with Hamiltonian
\begin{equation}
H(x,\theta)=\frac{mg}{8\pi\eta}\frac{1}{\sqrt{x^{2}+z^{2}}}+2v_{A}\cos\theta
\end{equation}
which has the dimension of velocity and is a constant of motion \citep{divfreeH}. Position
and angle are canonically conjugate variables and the dynamics preserves
the two- form $dx\wedge d\theta$ \citep{arnol2013mathematical}. Level sets $H(x,\theta)=E$ of
the Hamiltonian, shown in Fig.(\ref{fig:unboundedPlots}a), define
orbits in the $x-\theta$ plane labelled by the ``energy'' $E$.
For closed orbits, $\theta$ vanishes at the turning points and $x$
reaches its maximum $x_{m}$, giving $E=mg/8\pi\eta\sqrt{x_{m}^{2}+z^{2}}+2v_{A}\geq2v_{A}$
as a bound for such orbits. Trajectories on the orbit are obtained
by integrating $dt=-dx/\partial_{\theta}H=d\theta/\partial_{x}H$
at constant energy, from which the period follows directly. For small
oscillations, a quadratic approximation to the Hamiltonian shows that
$x$ and $\theta$ vary harmonically with frequency $\omega_{0}=2\pi/T_{0}=\sqrt{mgv_{A}/\eta z^{3}}$.
For large oscillations, the trajectory integrals can be obtained exactly
in terms of elliptic functions \citep{Grad&Rhy}. The result for the
period $T_{E}$, scaled by the frequency of small oscillations, is
shown in Fig.(\ref{fig:unboundedPlots}b). The mean height is driven by
the Hamiltonian limit cycle and its change per period is
\begin{equation}
\frac{\Delta h}{T_{E}}=-\left(v_{0}+E\right)+\langle3v_{A}\cos\theta-v_{0}\frac{3bz^{2}}{4r^{3}}\rangle
\end{equation}
where angled brackets denote orbital averages at energy $E$ and
$v_{0}=\frac{mg}{6\pi\eta b}$. The right hand side averages can be obtained
exactly in terms of elliptic functions \citep{Grad&Rhy}. The mean sedimentation
speed $\Delta h/T_{E}$ thus obtained is shown in Fig.(\ref{fig:unboundedPlots}c).
The root of the above equation determines the critical value $E_{0}$
of the energy above (below) which the net vertical motion is upward
(downward). A typical sedimenting trajectory, $E>E_{0}$, is shown
in $x-\theta-h$ space in Fig.(\ref{fig:unboundedPlots}d).

The symplectic structure is destroyed when the re-orienting effect
of the gravitational torque is included. The Hamiltonian increases
monotonically at the rate $\dot{H}=T_{R}v_{A}\sin^{2}\theta/4\pi\eta b^{3}$
to its maximum value of $mg/8\pi\eta z+2v_{A}$ at $x=\theta=0$,
and this corresponds to the pair sedimenting with a vertical separation
$z$ and oriented vertically. We next examine how reflected flow alters
these exact results.

\begin{figure}
\includegraphics[width=1\columnwidth]{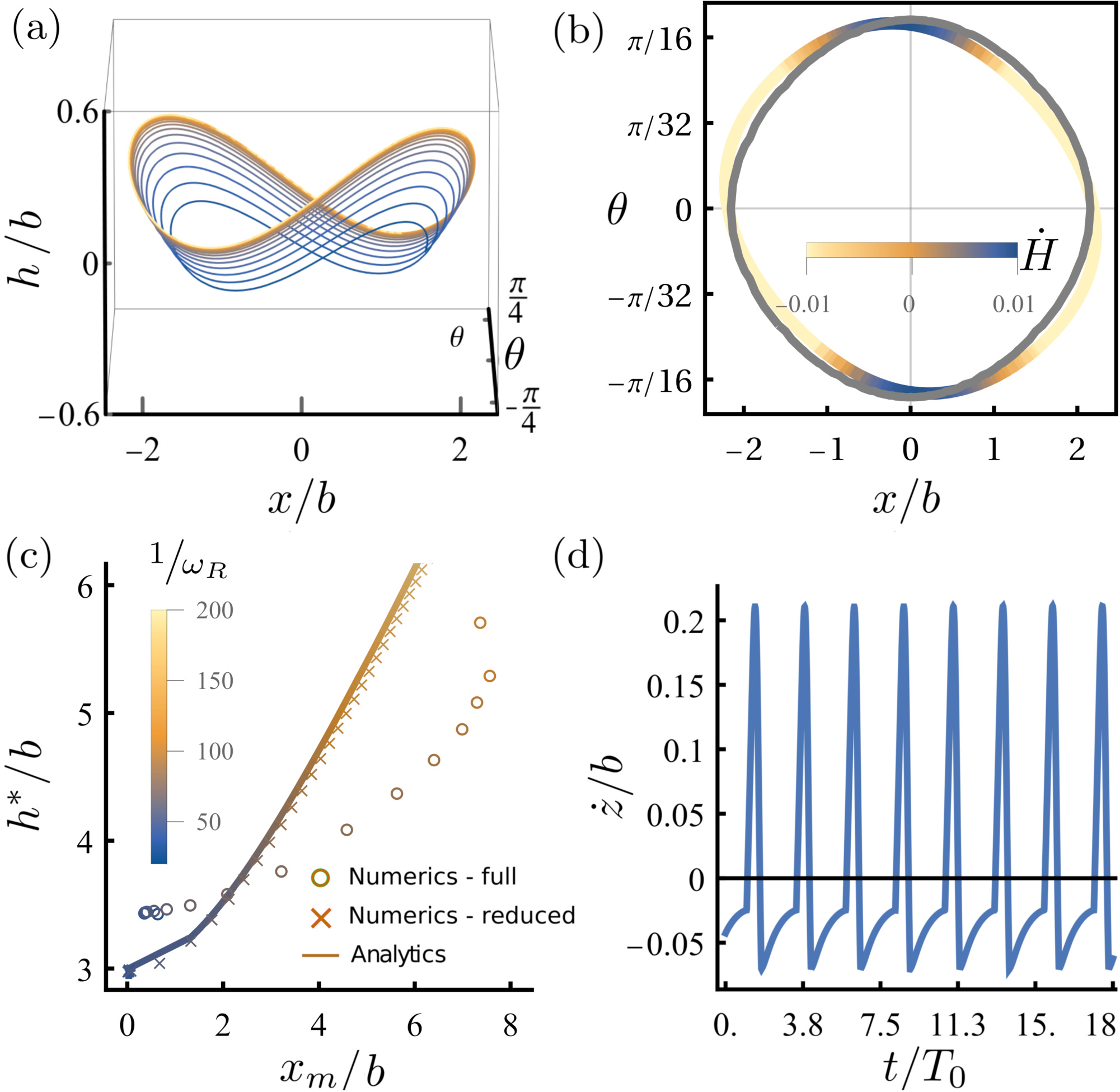}

\caption{\label{fig:perturbationpic}Perturbative solution of the two-body
system near the bottom plane, (a) a stable limit cycle is formed in
the presence of the bottom plane at height $h^{\star}$ and ``energy'' $E^{\star}$, (b) an overlay of the numeric and
analytic contours with $E=2.2$. The colormap shows instantaneous
``power'' input into the numerical limit cycle from perturbative
effects which produce small deformations to the shape of the orbit,
(c) the period averaged levitation height from the bottom plane
$h^{\star}/b$ against the maximum amplitude $x_{m}$ for the limit cycle
as $\omega_{R}$ is varied. Excellent quantitative agreement between the
analytic solution in Eqs.(\ref{eq:sedCOndition}-\ref{eq:Hcondition})
and the reduced numerics justifies, \textit{a posteriori}, averaging over the Hamiltonian orbit. Qualitative agreement between analytics and the full equations of motion shows the monotonic increase of $h^{\star}$ with $E^{\star}$ survives when the complete dynamics is considered. A fixed value of $z/b=2$ was used for analytic and reduced numerical computations while a more realistic short-ranged repulsive harmonic potential was implemented in the full numerics, resulting in a non-zero $\dot{z}$.  (d) the spikes in $\dot{z}/b$ are caused by steric repulsion induced by the short-ranged potential. Thus, over an oscillation, the value of $z$ changes, leading to quantitative disagreement as seen in (c).}\vspace{-.6cm}
\end{figure}

\emph{Limit cycle near bottom plane}: The effect of reflected flow
appears at different orders of $h$ in the dynamical system. In decreasing
order of importance, $h$ dynamics receive an $O(b/h)$ reduction
in the effective mobility, $z$ dynamics receive an $O(b^{2}/h^{2})$
hydrodynamic attraction, $x$ dynamics receive an $O(b^{2}/h^{2})$
hydrodynamic repulsion, and $\theta$ dynamics receive an $O(b^{3}/h^{3})$
contribution to reorientation. A levitating state at a mean height
$h^{\star}$ can exist if the change in mean height per period
is zero, giving
\begin{equation}
-\left(v_{0}+E\right)+\langle3v_{A}\cos\theta-v_{0}\left(\frac{3bz^{2}}{4r^{3}}-\frac{3b}{2h}\right)\rangle=0.\label{eq:sedCOndition}
\end{equation}
The rate of change of the Hamiltonian on the true limit cycle is $\dot{H}=T_{R}v_{A}\sin^{2}\theta/4\pi\eta b^{3}-\left(\frac{mg}{8\pi\eta}\right)^{2}\frac{x^{2}}{2\left(x^{2}+z^{2}\right)^{3/2}h^{2}}$
and, if this is to vanish over an orbit, we must have
\begin{equation}
\langle\dot{H}\rangle=\langle2\omega_{R}v_{A}\sin^{2}\theta-\frac{9b^{2}v_{0}^{2}x^{2}}{32\left(x^{2}+z^{2}\right)^{3/2}h^{2}}\rangle=0.\label{eq:Hcondition}
\end{equation}
To $O(z^{3}/h^{3})$ the average over the true limit cycle can be
replaced by an average over the Hamiltonian limit cycle at some energy
$E^{\star}$ \citep{krylov1947}. The above pair of equations, in
which averages are taken over Hamiltonian orbits, implicitly determines
unique values of $h^{\star}$ and $E^{\star}$ which define the levitating,
periodic, stable limit cycle in the presence of reflected flow (see Fig.(\ref{fig:perturbationpic})). Unlike  \citep{drescher2009}, we do not find a Hopf bifurcation but rather a transient decay of the Hamiltonian limit cycle into a stable one \citep{noteDrescher}.

\emph{Fixed point at top plane}: For energy values $E<E_{0}$ the
net vertical motion is upwards. In this case, the dynamical system must be modified to account for the
proximity of the top plane. This is obtained by replacing
$h$ by $\pm(L-h)$ in Eqs.(\ref{eq:dynamicalSystem-1}-\ref{eq:dynamicalSystem}).
The effect of reflected flow from the top surface is now the reverse
and instead of being destabilising is stabilising. The limit cycle
is destroyed and, instead, a dimerised state is obtained due to the
attractive flow of the monopoles pointing away from the plane \citep{squiresWall}. This
is identical in mechanism but distinct in detail to flow-induced phase
separation of active particles which swim into the plane surface \citep{singh2016crystallization}.

\emph{Conclusion: } We have presented overdamped equations for the
hydrodynamically interacting dynamics of a pair of self-propelling, self-spinning particles
in the presence of external forces and torques confined between planes.
We have identified a regime away from both planes where the
dynamics is symplectic, with a Hamiltonian determing periodic orbits. A second regime
near the bottom plane has a limit cycle which can be related perturbatively to the Hamiltonian
oscillator. A third regime at the top plane is non-periodic with a fixed point. Qualitatively,
the reflection of the monopolar flow at the top/bottom plane approximates extensile/contractile
dipolar flow \citep{squiresWall} and destablises/stabilises the Hamiltonian limit cycle.  A simple criterion
has been found for bifurcations between these states, determined by the value of Hamiltonian
$E$. Notably, this mechanism is operative at both no-slip and no-shear planes as it appears
at leading order in the reflection; next-to-leading order terms only alter the time-scales of motion
(sec IV of SI \citep{siText}). Our theory can be extended  to many-particle levitating states of active matter, such as active emulsion droplets, by including  next to leading order effects from symmetric dipoles and will be presented elsewhere \citep{bolitho}. Our work shows that active matter, which breaks time-reversal invariance \citep{HowFarEqActiveMatter} and is inherently dissipative \citep{singh2018generalized}, may nonetheless be described by Hamiltonian dynamics.

\emph{Acknowledgements:} We acknowledge the EPSRC (AB), the Royal
Society-SERB Newton International Fellowship (RS) and the Isaac Newton
Trust (RA) for support. We thank Prof. M. E. Cates for critical remarks
and Prof. R. E. Goldstein for helpful discussions and for bringing \citep{squiresWall} to our attention.We note that, since submission of this manuscript, a detailed state-of-the-art boundary element analysis of a pair of Volvox has appeared in \citep{ishikawa2020stability}.

\begin{thebibliography}{36}%
	\makeatletter
	\providecommand \@ifxundefined [1]{%
		\@ifx{#1\undefined}
	}%
	\providecommand \@ifnum [1]{%
		\ifnum #1\expandafter \@firstoftwo
		\else \expandafter \@secondoftwo
		\fi
	}%
	\providecommand \@ifx [1]{%
		\ifx #1\expandafter \@firstoftwo
		\else \expandafter \@secondoftwo
		\fi
	}%
	\providecommand \natexlab [1]{#1}%
	\providecommand \enquote  [1]{``#1''}%
	\providecommand \bibnamefont  [1]{#1}%
	\providecommand \bibfnamefont [1]{#1}%
	\providecommand \citenamefont [1]{#1}%
	\providecommand \href@noop [0]{\@secondoftwo}%
	\providecommand \href [0]{\begingroup \@sanitize@url \@href}%
	\providecommand \@href[1]{\@@startlink{#1}\@@href}%
	\providecommand \@@href[1]{\endgroup#1\@@endlink}%
	\providecommand \@sanitize@url [0]{\catcode `\\12\catcode `\$12\catcode
		`\&12\catcode `\#12\catcode `\^12\catcode `\_12\catcode `\%12\relax}%
	\providecommand \@@startlink[1]{}%
	\providecommand \@@endlink[0]{}%
	\providecommand \url  [0]{\begingroup\@sanitize@url \@url }%
	\providecommand \@url [1]{\endgroup\@href {#1}{\urlprefix }}%
	\providecommand \urlprefix  [0]{URL }%
	\providecommand \Eprint [0]{\href }%
	\providecommand \doibase [0]{http://dx.doi.org/}%
	\providecommand \selectlanguage [0]{\@gobble}%
	\providecommand \bibinfo  [0]{\@secondoftwo}%
	\providecommand \bibfield  [0]{\@secondoftwo}%
	\providecommand \translation [1]{[#1]}%
	\providecommand \BibitemOpen [0]{}%
	\providecommand \bibitemStop [0]{}%
	\providecommand \bibitemNoStop [0]{.\EOS\space}%
	\providecommand \EOS [0]{\spacefactor3000\relax}%
	\providecommand \BibitemShut  [1]{\csname bibitem#1\endcsname}%
	\let\auto@bib@innerbib\@empty
	\bibitem [{\citenamefont {Lighthill}(1952)}]{lighthill1952}%
	\BibitemOpen
	\bibfield  {author} {\bibinfo {author} {\bibfnamefont {M.~J.}\ \bibnamefont
			{Lighthill}},\ }\bibfield  {title} {\enquote {\bibinfo {title} {{On the
					squirming motion of nearly spherical deformable bodies through liquids at
					very small {R}eynolds numbers}},}\ }\href {\doibase 10.1002/cpa.3160050201}
	{\bibfield  {journal} {\bibinfo  {journal} {Commun. Pure. Appl. Math.}\
		}\textbf {\bibinfo {volume} {5}},\ \bibinfo {pages} {109--118} (\bibinfo
		{year} {1952})}\BibitemShut {NoStop}%
	\bibitem [{\citenamefont {Blake}(1971{\natexlab{a}})}]{blake1971a}%
	\BibitemOpen
	\bibfield  {author} {\bibinfo {author} {\bibfnamefont {J.~R.}\ \bibnamefont
			{Blake}},\ }\bibfield  {title} {\enquote {\bibinfo {title} {{A spherical
					envelope approach to ciliary propulsion}},}\ }\href {\doibase
		10.1017/S002211207100048X} {\bibfield  {journal} {\bibinfo  {journal} {J.
				Fluid Mech.}\ }\textbf {\bibinfo {volume} {46}},\ \bibinfo {pages} {199--208}
		(\bibinfo {year} {1971}{\natexlab{a}})}\BibitemShut {NoStop}%
	\bibitem [{\citenamefont {Anderson}(1989)}]{anderson1989colloid}%
	\BibitemOpen
	\bibfield  {author} {\bibinfo {author} {\bibfnamefont {J.~L.}\ \bibnamefont
			{Anderson}},\ }\bibfield  {title} {\enquote {\bibinfo {title} {{Colloid
					transport by interfacial forces}},}\ }\href {\doibase
		10.1146/annurev.fl.21.010189.000425} {\bibfield  {journal} {\bibinfo
			{journal} {Annu. Rev. Fluid Mech.}\ }\textbf {\bibinfo {volume} {21}},\
		\bibinfo {pages} {61--99} (\bibinfo {year} {1989})}\BibitemShut {NoStop}%
	\bibitem [{\citenamefont {Lauga}\ and\ \citenamefont
		{Powers}(2009)}]{lauga2009}%
	\BibitemOpen
	\bibfield  {author} {\bibinfo {author} {\bibfnamefont {E.}~\bibnamefont
			{Lauga}}\ and\ \bibinfo {author} {\bibfnamefont {T.~R.}\ \bibnamefont
			{Powers}},\ }\bibfield  {title} {\enquote {\bibinfo {title} {{The
					hydrodynamics of swimming microorganisms}},}\ }\href {\doibase
		10.1088/0034-4885/72/9/096601} {\bibfield  {journal} {\bibinfo  {journal}
			{Rep. Prog. Phys.}\ }\textbf {\bibinfo {volume} {72}},\ \bibinfo {pages}
		{096601} (\bibinfo {year} {2009})}\BibitemShut {NoStop}%
	\bibitem [{\citenamefont {Drescher}\ \emph {et~al.}(2010)\citenamefont
		{Drescher}, \citenamefont {Goldstein}, \citenamefont {Michel}, \citenamefont
		{Polin},\ and\ \citenamefont {Tuval}}]{drescher2010}%
	\BibitemOpen
	\bibfield  {author} {\bibinfo {author} {\bibfnamefont {K.}~\bibnamefont
			{Drescher}}, \bibinfo {author} {\bibfnamefont {R.~E.}\ \bibnamefont
			{Goldstein}}, \bibinfo {author} {\bibfnamefont {N.}~\bibnamefont {Michel}},
		\bibinfo {author} {\bibfnamefont {M.}~\bibnamefont {Polin}}, \ and\ \bibinfo
		{author} {\bibfnamefont {I.}~\bibnamefont {Tuval}},\ }\bibfield  {title}
	{\enquote {\bibinfo {title} {{Direct measurement of the flow field around
					swimming microorganisms}},}\ }\href {\doibase 10.1103/PhysRevLett.105.168101}
	{\bibfield  {journal} {\bibinfo  {journal} {Phys. Rev. Lett.}\ }\textbf
		{\bibinfo {volume} {105}},\ \bibinfo {pages} {168101} (\bibinfo {year}
		{2010})}\BibitemShut {NoStop}%
	\bibitem [{\citenamefont {Koch}\ and\ \citenamefont
		{Subramanian}(2011)}]{koch2011}%
	\BibitemOpen
	\bibfield  {author} {\bibinfo {author} {\bibfnamefont {D.~L.}\ \bibnamefont
			{Koch}}\ and\ \bibinfo {author} {\bibfnamefont {G.}~\bibnamefont
			{Subramanian}},\ }\bibfield  {title} {\enquote {\bibinfo {title} {{Collective
					hydrodynamics of swimming microorganisms: Living fluids}},}\ }\href {\doibase
		10.1146/annurev-fluid-121108-145434} {\bibfield  {journal} {\bibinfo
			{journal} {Annu. Rev. Fluid Mech.}\ }\textbf {\bibinfo {volume} {43}},\
		\bibinfo {pages} {637--659} (\bibinfo {year} {2011})}\BibitemShut {NoStop}%
	\bibitem [{\citenamefont {Marchetti}\ \emph {et~al.}(2013)\citenamefont
		{Marchetti}, \citenamefont {Joanny}, \citenamefont {Ramaswamy}, \citenamefont
		{Liverpool}, \citenamefont {Prost}, \citenamefont {Rao},\ and\ \citenamefont
		{Simha}}]{marchetti2013}%
	\BibitemOpen
	\bibfield  {author} {\bibinfo {author} {\bibfnamefont {M.~C.}\ \bibnamefont
			{Marchetti}}, \bibinfo {author} {\bibfnamefont {J.~F.}\ \bibnamefont
			{Joanny}}, \bibinfo {author} {\bibfnamefont {S.}~\bibnamefont {Ramaswamy}},
		\bibinfo {author} {\bibfnamefont {T.~B.}\ \bibnamefont {Liverpool}}, \bibinfo
		{author} {\bibfnamefont {J.}~\bibnamefont {Prost}}, \bibinfo {author}
		{\bibfnamefont {Madan}\ \bibnamefont {Rao}}, \ and\ \bibinfo {author}
		{\bibfnamefont {R.~A.}\ \bibnamefont {Simha}},\ }\bibfield  {title}
	{\enquote {\bibinfo {title} {{Hydrodynamics of soft active matter}},}\ }\href
	{\doibase 10.1103/RevModPhys.85.1143} {\bibfield  {journal} {\bibinfo
			{journal} {Rev. Mod. Phys.}\ }\textbf {\bibinfo {volume} {85}},\ \bibinfo
		{pages} {1143--1189} (\bibinfo {year} {2013})}\BibitemShut {NoStop}%
	\bibitem [{\citenamefont {Ramaswamy}(2017)}]{ramaswamy2017active}%
	\BibitemOpen
	\bibfield  {author} {\bibinfo {author} {\bibfnamefont {S.}~\bibnamefont
			{Ramaswamy}},\ }\bibfield  {title} {\enquote {\bibinfo {title} {{Active
					matter}},}\ }\href
	{http://iopscience.iop.org/article/10.1088/1742-5468/aa6bc5/meta} {\bibfield
		{journal} {\bibinfo  {journal} {J. Stat. Mech.}\ }\textbf {\bibinfo {volume}
			{2017}},\ \bibinfo {pages} {054002} (\bibinfo {year} {2017})}\BibitemShut
	{NoStop}%
	\bibitem [{\citenamefont {Ghose}\ and\ \citenamefont
		{Adhikari}(2014)}]{Ghose2014}%
	\BibitemOpen
	\bibfield  {author} {\bibinfo {author} {\bibfnamefont {S.}~\bibnamefont
			{Ghose}}\ and\ \bibinfo {author} {\bibfnamefont {R.}~\bibnamefont
			{Adhikari}},\ }\bibfield  {title} {\enquote {\bibinfo {title} {{Irreducible
					Representations of Oscillatory and Swirling Flows in Active Soft Matter}},}\
	}\href {\doibase 10.1103/PhysRevLett.112.118102} {\bibfield  {journal}
		{\bibinfo  {journal} {Phys. Rev. Lett.}\ }\textbf {\bibinfo {volume} {112}},\
		\bibinfo {pages} {118102} (\bibinfo {year} {2014})}\BibitemShut {NoStop}%
	\bibitem [{\citenamefont {Pedley}\ \emph {et~al.}(2016)\citenamefont {Pedley},
		\citenamefont {Brumley},\ and\ \citenamefont {Goldstein}}]{Pedley2016}%
	\BibitemOpen
	\bibfield  {author} {\bibinfo {author} {\bibfnamefont {T.~J.}\ \bibnamefont
			{Pedley}}, \bibinfo {author} {\bibfnamefont {D.~R.}\ \bibnamefont {Brumley}},
		\ and\ \bibinfo {author} {\bibfnamefont {R.~E.}\ \bibnamefont {Goldstein}},\
	}\bibfield  {title} {\enquote {\bibinfo {title} {{Squirmers with swirl: a
					model for Volvox swimming}},}\ }\href {\doibase 10.1017/jfm.2016.306}
	{\bibfield  {journal} {\bibinfo  {journal} {Journal of Fluid Mechanics}\
		}\textbf {\bibinfo {volume} {798}},\ \bibinfo {pages} {165--186} (\bibinfo
		{year} {2016})}\BibitemShut {NoStop}%
	\bibitem [{\citenamefont {Drescher}\ \emph {et~al.}(2009)\citenamefont
		{Drescher}, \citenamefont {Leptos}, \citenamefont {Tuval}, \citenamefont
		{Ishikawa}, \citenamefont {Pedley},\ and\ \citenamefont
		{Goldstein}}]{drescher2009}%
	\BibitemOpen
	\bibfield  {author} {\bibinfo {author} {\bibfnamefont {K.}~\bibnamefont
			{Drescher}}, \bibinfo {author} {\bibfnamefont {K.~C.}\ \bibnamefont
			{Leptos}}, \bibinfo {author} {\bibfnamefont {I.}~\bibnamefont {Tuval}},
		\bibinfo {author} {\bibfnamefont {T.}~\bibnamefont {Ishikawa}}, \bibinfo
		{author} {\bibfnamefont {T.~J.}\ \bibnamefont {Pedley}}, \ and\ \bibinfo
		{author} {\bibfnamefont {R.~E.}\ \bibnamefont {Goldstein}},\ }\bibfield
	{title} {\enquote {\bibinfo {title} {{Dancing Volvox: Hydrodynamic bound
					states of swimming algae}},}\ }\href {\doibase
		10.1103/PhysRevLett.102.168101} {\bibfield  {journal} {\bibinfo  {journal}
			{Phys. Rev. Lett.}\ }\textbf {\bibinfo {volume} {102}},\ \bibinfo {pages}
		{168101} (\bibinfo {year} {2009})}\BibitemShut {NoStop}%
	\bibitem [{\citenamefont {Goldstein}(2015)}]{goldstein2015green}%
	\BibitemOpen
	\bibfield  {author} {\bibinfo {author} {\bibfnamefont {R.~E.}\ \bibnamefont
			{Goldstein}},\ }\bibfield  {title} {\enquote {\bibinfo {title} {{Green algae
					as model organisms for biological fluid dynamics}},}\ }\href {\doibase
		10.1146/annurev-fluid-010313-141426} {\bibfield  {journal} {\bibinfo
			{journal} {Ann. Rev. Fluid Mech.}\ }\textbf {\bibinfo {volume} {47}},\
		\bibinfo {pages} {343--375} (\bibinfo {year} {2015})}\BibitemShut {NoStop}%
	\bibitem [{\citenamefont {Palacci}\ \emph {et~al.}(2013)\citenamefont
		{Palacci}, \citenamefont {Sacanna}, \citenamefont {Steinberg}, \citenamefont
		{Pine},\ and\ \citenamefont {Chaikin}}]{palacci2013living}%
	\BibitemOpen
	\bibfield  {author} {\bibinfo {author} {\bibfnamefont {J.}~\bibnamefont
			{Palacci}}, \bibinfo {author} {\bibfnamefont {S.}~\bibnamefont {Sacanna}},
		\bibinfo {author} {\bibfnamefont {A.~P.}\ \bibnamefont {Steinberg}}, \bibinfo
		{author} {\bibfnamefont {D.~J.}\ \bibnamefont {Pine}}, \ and\ \bibinfo
		{author} {\bibfnamefont {P.~M.}\ \bibnamefont {Chaikin}},\ }\bibfield
	{title} {\enquote {\bibinfo {title} {{Living crystals of light-activated
					colloidal surfers}},}\ }\href {\doibase 10.1126/science.1230020} {\bibfield
		{journal} {\bibinfo  {journal} {Science}\ }\textbf {\bibinfo {volume}
			{339}},\ \bibinfo {pages} {936--940} (\bibinfo {year} {2013})}\BibitemShut
	{NoStop}%
	\bibitem [{\citenamefont {Palacci}\ \emph {et~al.}(2010)\citenamefont
		{Palacci}, \citenamefont {Cottin-Bizonne}, \citenamefont {Ybert},\ and\
		\citenamefont {Bocquet}}]{palacci2010sedimentation}%
	\BibitemOpen
	\bibfield  {author} {\bibinfo {author} {\bibfnamefont {J.}~\bibnamefont
			{Palacci}}, \bibinfo {author} {\bibfnamefont {C.}~\bibnamefont
			{Cottin-Bizonne}}, \bibinfo {author} {\bibfnamefont {C.}~\bibnamefont
			{Ybert}}, \ and\ \bibinfo {author} {\bibfnamefont {L.}~\bibnamefont
			{Bocquet}},\ }\bibfield  {title} {\enquote {\bibinfo {title} {{Sedimentation
					and Effective Temperature of Active Colloidal Suspensions}},}\ }\href
	{\doibase 10.1103/PhysRevLett.105.088304} {\bibfield  {journal} {\bibinfo
			{journal} {Phys. Rev. Lett.}\ }\textbf {\bibinfo {volume} {105}},\ \bibinfo
		{pages} {088304} (\bibinfo {year} {2010})}\BibitemShut {NoStop}%
	\bibitem [{\citenamefont {Buttinoni}\ \emph {et~al.}(2013)\citenamefont
		{Buttinoni}, \citenamefont {Bialk{\'e}}, \citenamefont {K{\"u}mmel},
		\citenamefont {L{\"o}wen}, \citenamefont {Bechinger},\ and\ \citenamefont
		{Speck}}]{buttinoni2013DynamicClustering}%
	\BibitemOpen
	\bibfield  {author} {\bibinfo {author} {\bibfnamefont {I.}~\bibnamefont
			{Buttinoni}}, \bibinfo {author} {\bibfnamefont {J.}~\bibnamefont
			{Bialk{\'e}}}, \bibinfo {author} {\bibfnamefont {F.}~\bibnamefont
			{K{\"u}mmel}}, \bibinfo {author} {\bibfnamefont {H.}~\bibnamefont
			{L{\"o}wen}}, \bibinfo {author} {\bibfnamefont {C.}~\bibnamefont
			{Bechinger}}, \ and\ \bibinfo {author} {\bibfnamefont {T.}~\bibnamefont
			{Speck}},\ }\bibfield  {title} {\enquote {\bibinfo {title} {{Dynamical
					Clustering and Phase Separation in Suspensions of Self-Propelled Colloidal
					Particles}},}\ }\href {\doibase 10.1103/PhysRevLett.110.238301} {\bibfield
		{journal} {\bibinfo  {journal} {Phys. Rev. Lett.}\ }\textbf {\bibinfo
			{volume} {110}},\ \bibinfo {pages} {238301} (\bibinfo {year}
		{2013})}\BibitemShut {NoStop}%
	\bibitem [{\citenamefont {Ebbens}\ and\ \citenamefont
		{Howse}(2010)}]{ebbens2010pursuit}%
	\BibitemOpen
	\bibfield  {author} {\bibinfo {author} {\bibfnamefont {S.~J.}\ \bibnamefont
			{Ebbens}}\ and\ \bibinfo {author} {\bibfnamefont {J.~R.}\ \bibnamefont
			{Howse}},\ }\bibfield  {title} {\enquote {\bibinfo {title} {{In pursuit of
					propulsion at the nanoscale}},}\ }\href {\doibase 10.1039/B918598D}
	{\bibfield  {journal} {\bibinfo  {journal} {Soft Matter}\ }\textbf {\bibinfo
			{volume} {6}},\ \bibinfo {pages} {726--738} (\bibinfo {year}
		{2010})}\BibitemShut {NoStop}%
	\bibitem [{\citenamefont {Smoluchowski}(1911)}]{smoluchowski1911mutual}%
	\BibitemOpen
	\bibfield  {author} {\bibinfo {author} {\bibfnamefont {M.}~\bibnamefont
			{Smoluchowski}},\ }\bibfield  {title} {\enquote {\bibinfo {title} {{On the
					mutual action of spheres which move in a viscous liquid}},}\ }\href@noop {}
	{\bibfield  {journal} {\bibinfo  {journal} {Bull. Acad. Sci. Cracovie A}\
		}\textbf {\bibinfo {volume} {1}},\ \bibinfo {pages} {28--39} (\bibinfo {year}
		{1911})}\BibitemShut {NoStop}%
	\bibitem [{\citenamefont {Brenner}(1963)}]{brenner1963}%
	\BibitemOpen
	\bibfield  {author} {\bibinfo {author} {\bibfnamefont {H.}~\bibnamefont
			{Brenner}},\ }\bibfield  {title} {\enquote {\bibinfo {title} {{The {S}tokes
					resistance of an arbitrary particle}},}\ }\href {\doibase
		10.1016/0009-2509(64)85084-3} {\bibfield  {journal} {\bibinfo  {journal}
			{Chem. Engg. Sci.}\ }\textbf {\bibinfo {volume} {18}},\ \bibinfo {pages}
		{1--25} (\bibinfo {year} {1963})}\BibitemShut {NoStop}%
	\bibitem [{\citenamefont {Ladd}(1988)}]{ladd1988}%
	\BibitemOpen
	\bibfield  {author} {\bibinfo {author} {\bibfnamefont {A.~J.~C.}\
			\bibnamefont {Ladd}},\ }\bibfield  {title} {\enquote {\bibinfo {title}New1
			{{Hydrodynamic interactions in a suspension of spherical particles}},}\
	}\href {\doibase 10.1063/1.454658} {\bibfield  {journal} {\bibinfo  {journal}
			{J. Chem. Phys.}\ }\textbf {\bibinfo {volume} {88}},\ \bibinfo {pages}
		{5051--5063} (\bibinfo {year} {1988})}\BibitemShut {NoStop}%
	\bibitem [{\citenamefont {Squires}\ and\ \citenamefont
		{Brenner}(2000)}]{squiresWall}%
	\BibitemOpen
	\bibfield  {author} {\bibinfo {author} {\bibfnamefont {T.~M.}\ \bibnamefont
			{Squires}}\ and\ \bibinfo {author} {\bibfnamefont {M.~P.}\ \bibnamefont
			{Brenner}},\ }\bibfield  {title} {\enquote {\bibinfo {title} {{Like-Charge
					Attraction and Hydrodynamic Interaction}},}\ }\href {\doibase
		10.1103/PhysRevLett.85.4976} {\bibfield  {journal} {\bibinfo  {journal}
			{Phys. Rev. Lett.}\ }\textbf {\bibinfo {volume} {85}},\ \bibinfo {pages}
		{4976--4979} (\bibinfo {year} {2000})}\BibitemShut {NoStop}%
	\bibitem [{\citenamefont {Squires}(2001)}]{squires2001effective}%
	\BibitemOpen
	\bibfield  {author} {\bibinfo {author} {\bibfnamefont {T.~M.}\ \bibnamefont
			{Squires}},\ }\bibfield  {title} {\enquote {\bibinfo {title} {{Effective
					pseudo-potentials of hydrodynamic origin}},}\ }\href {\doibase
		10.1017/S0022112001005432} {\bibfield  {journal} {\bibinfo  {journal} {J.
				Fluid Mech.}\ }\textbf {\bibinfo {volume} {443}},\ \bibinfo {pages}
		{403--412} (\bibinfo {year} {2001})}\BibitemShut {NoStop}%
	\bibitem [{\citenamefont {Singh}\ and\ \citenamefont
		{Adhikari}(2016)}]{singh2016crystallization}%
	\BibitemOpen
	\bibfield  {author} {\bibinfo {author} {\bibfnamefont {R.}~\bibnamefont
			{Singh}}\ and\ \bibinfo {author} {\bibfnamefont {R.}~\bibnamefont
			{Adhikari}},\ }\bibfield  {title} {\enquote {\bibinfo {title} {{Universal
					hydrodynamic mechanisms for crystallization in active colloidal
					suspensions}},}\ }\href {\doibase 10.1103/PhysRevLett.117.228002} {\bibfield
		{journal} {\bibinfo  {journal} {Phys. Rev. Lett.}\ }\textbf {\bibinfo
			{volume} {117}},\ \bibinfo {pages} {228002} (\bibinfo {year}
		{2016})}\BibitemShut {NoStop}%
	\bibitem [{\citenamefont {Nash}\ \emph {et~al.}(2010)\citenamefont {Nash},
		\citenamefont {Adhikari}, \citenamefont {Tailleur},\ and\ \citenamefont
		{Cates}}]{nash2010}%
	\BibitemOpen
	\bibfield  {author} {\bibinfo {author} {\bibfnamefont {R.~W.}\ \bibnamefont
			{Nash}}, \bibinfo {author} {\bibfnamefont {R.}~\bibnamefont {Adhikari}},
		\bibinfo {author} {\bibfnamefont {J.}~\bibnamefont {Tailleur}}, \ and\
		\bibinfo {author} {\bibfnamefont {M.~E.}\ \bibnamefont {Cates}},\ }\bibfield
	{title} {\enquote {\bibinfo {title} {{Run-and-Tumble Particles with
					Hydrodynamics: Sedimentation, Trapping, and Upstream Swimming}},}\ }\href
	{\doibase 10.1103/PhysRevLett.104.258101} {\bibfield  {journal} {\bibinfo
			{journal} {Phys. Rev. Lett.}\ }\textbf {\bibinfo {volume} {104}},\ \bibinfo
		{pages} {258101} (\bibinfo {year} {2010})}\BibitemShut {NoStop}%
	\bibitem [{\citenamefont {Singh}\ \emph {et~al.}(2015)\citenamefont {Singh},
		\citenamefont {Ghose},\ and\ \citenamefont {Adhikari}}]{singh2015many}%
	\BibitemOpen
	\bibfield  {author} {\bibinfo {author} {\bibfnamefont {R.}~\bibnamefont
			{Singh}}, \bibinfo {author} {\bibfnamefont {S.}~\bibnamefont {Ghose}}, \ and\
		\bibinfo {author} {\bibfnamefont {R.}~\bibnamefont {Adhikari}},\ }\bibfield
	{title} {\enquote {\bibinfo {title} {{Many-body microhydrodynamics of
					colloidal particles with active boundary layers}},}\ }\href
	{http://stacks.iop.org/1742-5468/2015/i=6/a=P06017} {\bibfield  {journal}
		{\bibinfo  {journal} {J. Stat. Mech}\ }\textbf {\bibinfo {volume} {2015}},\
		\bibinfo {pages} {P06017} (\bibinfo {year} {2015})}\BibitemShut {NoStop}%
	\bibitem{greenFuncApprox} 
	In this limit, the Green's function can be approximated by that of an unbounded or semi-unbounded geometry.
	\bibitem [{\citenamefont {Singh}\ and\ \citenamefont
		{Adhikari}(2018)}]{singh2018generalized}%
	\BibitemOpen
	\bibfield  {author} {\bibinfo {author} {\bibfnamefont {R.}~\bibnamefont
			{Singh}}\ and\ \bibinfo {author} {\bibfnamefont {R.}~\bibnamefont
			{Adhikari}},\ }\bibfield  {title} {\enquote {\bibinfo {title} {{Generalized
					{S}tokes laws for active colloids and their applications}},}\ }\href
	{\doibase 10.1088/2399-6528/aaab0d} {\bibfield  {journal} {\bibinfo
			{journal} {J. Phys. Commun.}\ }\textbf {\bibinfo {volume} {2}},\ \bibinfo
		{pages} {025025} (\bibinfo {year} {2018})}\BibitemShut {NoStop}%
	\bibitem [{siT()}]{siText}%
	\BibitemOpen
	\href@noop {} {\enquote {\bibinfo {title} {{See Supplemental Material at [to
					be inserted] which includes the details of the calculations, simulation
					details, and movies of dynamics.}}}\ }\BibitemShut {NoStop}%
	\bibitem [{\citenamefont {Pozrikidis}(1992)}]{pozrikidis1992}%
	\BibitemOpen
	\bibfield  {author} {\bibinfo {author} {\bibfnamefont {C.}~\bibnamefont
			{Pozrikidis}},\ }\href {\doibase 10.1017/CBO9780511624124} {\emph {\bibinfo
			{title} {{Boundary Integral and Singularity Methods for Linearized Viscous
					Flow}}}}\ (\bibinfo  {publisher} {Cambridge University Press},\ \bibinfo
	{year} {1992})\BibitemShut {NoStop}%
	\bibitem [{\citenamefont {Pedley}\ and\ \citenamefont
		{Kessler}(1992)}]{pedley1992}%
	\BibitemOpen
	\bibfield  {author} {\bibinfo {author} {\bibfnamefont {T.~J.}\ \bibnamefont
			{Pedley}}\ and\ \bibinfo {author} {\bibfnamefont {J.~O.}\ \bibnamefont
			{Kessler}},\ }\bibfield  {title} {\enquote {\bibinfo {title} {{Hydrodynamic
					phenomena in suspensions of swimming microorganisms}},}\ }\href@noop {}
	{\bibfield  {journal} {\bibinfo  {journal} {Annu. Rev. Fluid Mech.}\ }\textbf
		{\bibinfo {volume} {24}},\ \bibinfo {pages} {313--358} (\bibinfo {year}
		{1992})}\BibitemShut {NoStop}%
		\bibitem{noteStability} 
	Gravitational torque therefore also provides robustness to the dynamics against out-of-plane perturbations.
	\bibitem{divfreeH} 
	Any divergence-free vector field in 2-dimensions is a Hamiltonian flow \citep{arnol2013mathematical}.
	\bibitem [{\citenamefont {Arnold}(2013)}]{arnol2013mathematical}%
	\BibitemOpen
	\bibfield  {author} {\bibinfo {author} {\bibfnamefont {V.~I.}\ \bibnamefont
			{Arnold}},\ }\href@noop {} {\emph {\bibinfo {title} {{Mathematical methods of
					classical mechanics}}}},\ Vol.~\bibinfo {volume} {60}\ (\bibinfo  {publisher}
	{Springer, New York},\ \bibinfo {year} {2013})\BibitemShut {NoStop}%
	\bibitem [{\citenamefont {Gradshteyn}\ and\ \citenamefont
		{Ryzhik}(2014)}]{Grad&Rhy}%
	\BibitemOpen
	\bibfield  {author} {\bibinfo {author} {\bibfnamefont {I.S.}\ \bibnamefont
			{Gradshteyn}}\ and\ \bibinfo {author} {\bibfnamefont {I.M.}\ \bibnamefont
			{Ryzhik}},\ }\bibfield  {title} {\enquote {\bibinfo {title} {{Table of
					Integrals, Series, and Products}},}\ \ }(\bibinfo  {publisher} {Academic
		Press},\ \bibinfo {year} {2014})\ pp.\ \bibinfo {pages} {63--247}\BibitemShut
	{NoStop}%
	\bibitem [{\citenamefont {Krylov}\ and\ \citenamefont
		{Bogoli︠u︡bov}(1947)}]{krylov1947}%
	\BibitemOpen
	\bibfield  {author} {\bibinfo {author} {\bibfnamefont {N.M.}\ \bibnamefont
			{Krylov}}\ and\ \bibinfo {author} {\bibfnamefont {N.~N.}\ \bibnamefont
			{Bogoli︠u︡bov}},\ }\href@noop {} {\emph {\bibinfo {title} {{Introduction
					to non-linear mechanics}}}}\ (\bibinfo  {publisher} {Princeton Univ. Press},\
	\bibinfo {year} {1947})\BibitemShut {NoStop}%
		\bibitem{noteDrescher} 
	The reason for this discrepancy is that  \citep{drescher2009} do not consider the height to be dynamical and fix it, by fiat, to a constant value, around which linear stability is examined.
	\bibitem [{\citenamefont {$Al.$}()}]{bolitho}%
	\BibitemOpen
	\bibfield  {author} {\bibinfo {author} {\bibfnamefont {A.~Bolitho~$Et$.}\
			\bibnamefont {$Al.$}},\ }\bibfield  {title} {\enquote {\bibinfo {title}
			{{Flow-induced bound states of active particles}},}\ }\href@noop {} {\bibinfo
		{journal} {In preparation}\ }\BibitemShut {NoStop}%
	\bibitem [{\citenamefont {Fodor}\ \emph {et~al.}(2016)\citenamefont {Fodor},
		\citenamefont {Nardini}, \citenamefont {Cates}, \citenamefont {Tailleur},
		\citenamefont {Visco},\ and\ \citenamefont {van
			Wijland}}]{HowFarEqActiveMatter}%
	\BibitemOpen
	\bibfield  {journal} {  }\bibfield  {author} {\bibinfo {author} {\bibfnamefont
			{{\'E}.}\ \bibnamefont {Fodor}}, \bibinfo {author} {\bibfnamefont
			{C.}\ \bibnamefont {Nardini}}, \bibinfo {author} {\bibfnamefont
			{M.~E.}\ \bibnamefont {Cates}}, \bibinfo {author} {\bibfnamefont
			{J.}\ \bibnamefont {Tailleur}}, \bibinfo {author} {\bibfnamefont {P.}\
			\bibnamefont {Visco}}, \ and\ \bibinfo {author} {\bibfnamefont
			{F.}\ \bibnamefont {van Wijland}},\ }\bibfield  {title}
	{\enquote {\bibinfo {title} {{How Far from Equilibrium Is Active Matter?}}}\
	}\href {\doibase 10.1103/PhysRevLett.117.038103} {\bibfield  {journal}
		{\bibinfo  {journal} {Phys. Rev. Lett.}\ }\textbf {\bibinfo {volume} {117}},\
		\bibinfo {pages} {038103} (\bibinfo {year} {2016})}\BibitemShut {NoStop}%
	\bibitem [{\citenamefont {{R. Singh}}\ and\ \citenamefont
		{Adhikari}(2019)}]{singh2019PyStokes}%
	\BibitemOpen
	\bibfield  {author} {\bibinfo {author} {\bibnamefont {{R. Singh}}}\ and\
		\bibinfo {author} {\bibfnamefont {R.}~\bibnamefont {Adhikari}},\ }\bibfield
	{title} {\enquote {\bibinfo {title} {{Hydrodynamic and phoretic interactions
					of active particles in Python}},}\ }\href {https://arxiv.org/abs/1910.00909}
	{\bibfield  {journal} {\bibinfo  {journal} {arXiv:1910.00909}\ } (\bibinfo
		{year} {2019})}\BibitemShut {NoStop}%
	\bibitem [{\citenamefont {Blake}(1971{\natexlab{b}})}]{blake1971c}%
	\BibitemOpen
	\bibfield  {author} {\bibinfo {author} {\bibfnamefont {J.~R.}\ \bibnamefont
			{Blake}},\ }\bibfield  {title} {\enquote {\bibinfo {title} {{A note on the
					image system for a {S}tokeslet in a no-slip boundary}},}\ }\href {\doibase
		10.1017/S0305004100049902} {\bibfield  {journal} {\bibinfo  {journal} {Proc.
				Camb. Phil. Soc.}\ }\textbf {\bibinfo {volume} {70}},\ \bibinfo {pages}
		{303--310} (\bibinfo {year} {1971}{\natexlab{b}})}\BibitemShut {NoStop}%
	\bibitem [{\citenamefont {Aderogba}\ and\ \citenamefont
		{Blake}(1978)}]{aderogba1978action}%
	\BibitemOpen
	\bibfield  {author} {\bibinfo {author} {\bibfnamefont {K.}~\bibnamefont
			{Aderogba}}\ and\ \bibinfo {author} {\bibfnamefont {J.~R.}\ \bibnamefont
			{Blake}},\ }\bibfield  {title} {\enquote {\bibinfo {title} {{Action of a
					force near the planar surface between semi-infinite immiscible liquids at
					very low {R}eynolds numbers}},}\ }\href {\doibase 10.1017/S0004972700008819}
	{\bibfield  {journal} {\bibinfo  {journal} {Bull. Australian Math. Soc.}\
		}\textbf {\bibinfo {volume} {19}},\ \bibinfo {pages} {309--318} (\bibinfo
		{year} {1978})}\BibitemShut {NoStop}%
	\bibitem [{\citenamefont {Ishikawa}\ \emph {et~al.}(2020)\citenamefont
		{Ishikawa}, \citenamefont {Pedley}, \citenamefont {Drescher},\ and\
		\citenamefont {Goldstein}}]{ishikawa2020stability}%
	\BibitemOpen
	\bibfield  {author} {\bibinfo {author} {\bibfnamefont {T.}\ \bibnamefont
			{Ishikawa}}, \bibinfo {author} {\bibfnamefont {T.~J.}\ \bibnamefont
			{Pedley}}, \bibinfo {author} {\bibfnamefont {K.}~\bibnamefont {Drescher}}, \
		and\ \bibinfo {author} {\bibfnamefont {Raymond~E.}\ \bibnamefont
			{Goldstein}},\ }\href@noop {} {\enquote {\bibinfo {title} {{Stability of
					dancing Volvox}},}\ } (\bibinfo {year} {2020}),\ \Eprint
	{http://arxiv.org/abs/2001.02825} {arXiv:2001.02825 [physics.flu-dyn]}
	\BibitemShut {NoStop}%
	\bibitem [{\citenamefont {Thutupalli}\ \emph {et~al.}(2018)\citenamefont
		{Thutupalli}, \citenamefont {Geyer}, \citenamefont {Singh}, \citenamefont
		{Adhikari},\ and\ \citenamefont {Stone}}]{thutupalli2018FIPS}%
	\BibitemOpen
	\bibfield  {author} {\bibinfo {author} {\bibfnamefont {S.}~\bibnamefont
			{Thutupalli}}, \bibinfo {author} {\bibfnamefont {D.}~\bibnamefont {Geyer}},
		\bibinfo {author} {\bibfnamefont {R.}~\bibnamefont {Singh}}, \bibinfo
		{author} {\bibfnamefont {R.}~\bibnamefont {Adhikari}}, \ and\ \bibinfo
		{author} {\bibfnamefont {H.~A.}\ \bibnamefont {Stone}},\ }\bibfield  {title}
	{\enquote {\bibinfo {title} {{Flow-induced phase separation of active
					particles is controlled by boundary conditions}},}\ }\href {\doibase
		10.1073/pnas.1718807115} {\bibfield  {journal} {\bibinfo  {journal} {Proc.
				Natl. Acad. Sci.}\ }\textbf {\bibinfo {volume} {115}},\ \bibinfo {pages}
		{5403--5408} (\bibinfo {year} {2018})}\BibitemShut {NoStop}%
\end{thebibliography}
%
\newpage\widetext

\section*{Supplemental Information (SI)}

\section{full equations of motion and numerical solution}

We consider a system of active colloids labeled as $i=1,\dots N$
of radius $b$ in an incompressible fluid of viscosity $\eta$. The
centre of mass of the $i$th colloid is denoted by $\boldsymbol{R}_{i}$,
while a unit vector $\boldsymbol{p}_{i}$ denotes its orientation.
The translational $\boldsymbol{V}_{i}$ and rotational velocity $\boldsymbol{\Omega}_{i}$
is given from the sum of all the forces and torques acting on the
colloids

\[
\begin{aligned}m\boldsymbol{\dot{V}}_{i}=-\boldsymbol{\gamma}_{ij}^{TT}\cdot(\boldsymbol{V}_{j}-v_{A}\boldsymbol{p}_{j})-\boldsymbol{\gamma}_{ij}^{TR}\cdot(\boldsymbol{\Omega}_{j}-\omega_{A}\boldsymbol{p}_{j})+\boldsymbol{F}_{i}^{B} & =0\\
I\boldsymbol{\dot{\Omega}}_{i}\boldsymbol{=}-\boldsymbol{\gamma}_{ij}^{RT}\cdot(\boldsymbol{V}_{j}-v_{A}\boldsymbol{p}_{j})-\boldsymbol{\gamma}_{ij}^{RR}\cdot(\boldsymbol{\Omega}_{j}-\omega_{A}\boldsymbol{p}_{j})+\boldsymbol{T}_{i}^{B} & =0
\end{aligned}
\]
Here $v_{A}$ (and $\omega_{A}$) is the self-propulsion translational
(rotational) speed of an isolated colloid, $\boldsymbol{\gamma}^{\alpha\beta}$,
for $(\alpha,\beta=T,R)$, are friction tensors \citep{ladd1988},
while $\boldsymbol{F}_{i}^{B}$ and $\boldsymbol{T}_{i}^{B}$ are
the body forces and torques on the $i$th colloid.

In the microhydrodynamic regime, as applicable to colloidal scale,
the inertia can be ignored, and the rigid body motion is then given
as\citep{ladd1988,singh2018generalized}\begin{subequations}\label{eq:RBM-1}
	\begin{align*}
	\boldsymbol{V}_{i}=\boldsymbol{\mu}_{ij}^{TT}\cdot\boldsymbol{F}_{j}^{B} & +\boldsymbol{\mu}_{ij}^{TR}\cdot\boldsymbol{T}_{j}^{B}+v_{A}\boldsymbol{p}_{i}\\
	\boldsymbol{\Omega}_{i}=\boldsymbol{\mu}_{ij}^{RT}\cdot\boldsymbol{F}_{j}^{B} & +\boldsymbol{\mu}_{ij}^{RR}\cdot\boldsymbol{T}_{j}^{B}+\omega_{A}\boldsymbol{p}_{i}
	\end{align*}
\end{subequations}Here $\boldsymbol{\mu}^{\alpha\beta}$, for $(\alpha,\beta=T,R)$,
are the mobility matrices \citep{ladd1988}.

\begin{figure}
	\includegraphics[width=0.9\columnwidth]{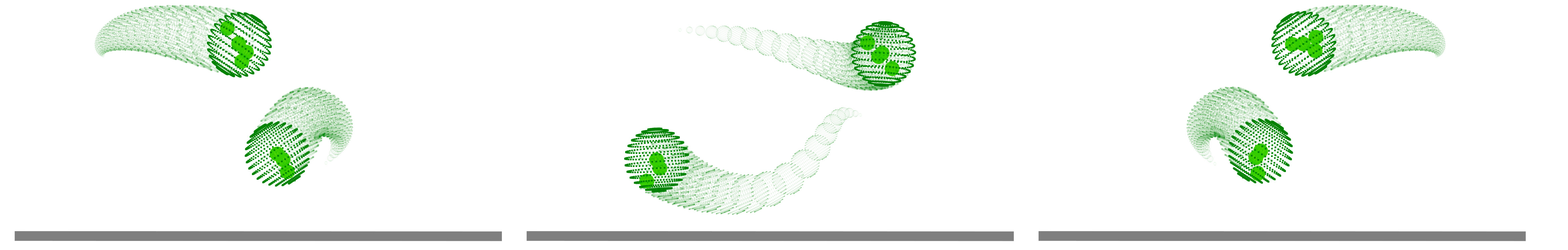}
	
	\caption{Stroboscopic images from the dynamics of two active particles near
		a plane no-shear surface. A similar dynamics is also obtained near
		a plane no-slip surface but with a longer time scale due to reduced
		strength of the hydrodynamic interactions. \label{fig:Limit-cycle-in}}
\end{figure}

The above equations have been simulated using PyStokes, a python package
for simulating Stokesian hydrodynamics \citep{singh2019PyStokes}.
The initial parameters were set to $b=1,\,v_{A}=1,\,v_{0}=1$. We
then study the system near a plane surface by computing the mobility
tensors using the appropriate Green's function of Stokes equation which
satisfies the boundary conditions of no-slip \citep{blake1971c} or
no-shear \citep{aderogba1978action} at a plane surface. Our system
of active particles near a plane surface has no periodic boundary
condition and the particles are allowed to explore the infinite half-space
around the surface. For simulations near the bottom plane, an additional
restoring torque of strength $\omega_{R}=0.022$ was added due to bottom-heaviness of the colloids. In this case, $z$ becomes a dynamic
variable and the separation changes greatly over the time period of
a cycle. In order to prevent the active particles getting too close
to one another an additional soft harmonic repulsion of strength $2$
was introduced when the particles came within $6.3$ units of radius
$b$ of one another. This kept the particles separated by an average
vertical distance $z=3$ during integration allowing comparison to
be made with the analytics and numerics of the reduced equations.
Making this potential soft and longer ranged made numerical integration
more stable and allowed larger integrator step sizes to be taken which
reduced the cost of running longer simulations. It was not necessary
to include a repulsive contact potential from the surface as the particles
were at least a distance $b$ away due to hydrodynamic repulsion
from the image charges. A two-particle simulation of above equation
leads to the formation of time-dependent bound state as described
in the main text. See Fig.(\ref{fig:Limit-cycle-in}) for snapshots
from the dynamics.
For simulations near the top surface, the same values were used.
\begin{table}[H]
	\renewcommand{\arraystretch}{2}\centering
	
	\begin{tabular}{|>{\centering}p{1.5cm}|>{\centering}p{5.5cm}|>{\centering}p{4.65cm}|>{\centering}p{5cm}|}
		\hline
		& $\boldsymbol{\mu}_{ij}^{TT}=(1+\frac{b^{2}}{3}\nabla^{2})\mathbf{G}$ & $\boldsymbol{\mu}_{ij}^{TR}=\frac{1}{2}\boldsymbol{\nabla}_{{\scriptscriptstyle \boldsymbol{R}_{j}}}\times\mathbf{G}$ & $\boldsymbol{\mu}_{ij}^{RR}=\frac{1}{4}\boldsymbol{\nabla}_{{\scriptscriptstyle \boldsymbol{R}_{i}}}\times\boldsymbol{\nabla}_{{\scriptscriptstyle \boldsymbol{R}_{j}}}\times\mathbf{G}$\tabularnewline
		\hline
		\hline
		Bottom\\plane & $\mu_{ii}^{zz}=\frac{1}{6\pi\eta b}\Big(1-\frac{3b}{4z_{i}}\Big)$. & $\tilde{\mu}_{ii}^{zy}=0$ & $\hat{\mu}_{ii}^{yy}=\frac{1}{8\pi\eta b^{3}}\Big(1-\frac{1}{16}\frac{b^{3}}{z_{i}^{3}}\Big)$\tabularnewline
		\hline
		Bottom\\plane & $\mu_{12}^{xz}=\frac{1}{8\pi\eta}\bigg[\frac{(x_{1}-x_{2})(z_{1}-z_{2})}{r^{3}}-\frac{(x_{1}-x_{2})(z_{1}+z_{2})}{r^{^{*}3}}\bigg]$ & $\tilde{\mu}_{12}^{xy}=\frac{1}{8\pi\eta}\bigg[\frac{(z_{1}-z_{2})}{r^{3}}-\frac{(z_{1}+z_{2})}{r^{^{*}3}}\bigg]$ & $\hat{\mu}_{12}^{yy}=-\frac{1}{16\pi\eta}\bigg[\frac{1}{r^{3}}-\frac{3(y_{1}-y_{2})(y_{1}-y_{2})}{r^{5}}+\frac{1}{r^{*3}}-\frac{3(y_{1}-y_{2})(y_{1}-y_{2})}{r^{^{*}5}}\bigg]$\tabularnewline
		\hline
		Bottom\\plane & $\mu_{12}^{zz}=\frac{1}{8\pi\eta}\bigg[\frac{1}{r}+\frac{(z_{1}-z_{2})(z_{1}-z_{2})}{r^{3}}-\frac{1}{r^{*}}-\frac{(z_{1}+z_{2})(z_{1}+z_{2})}{r^{^{*}3}}\bigg]$ & $\tilde{\mu}_{12}^{yz}=\frac{1}{8\pi\eta}\bigg[\frac{(x_{1}-x_{2})}{r^{3}}-\frac{(x_{1}-x_{2})}{r^{^{*}3}}\bigg]$ & $\hat{\mu}_{12}^{zy}=-\frac{1}{16\pi\eta}\bigg[\frac{1}{r^{3}}+\frac{3(y_{1}-y_{2})(z_{1}-z_{2})}{r^{5}}-\frac{1}{r^{*3}}-\frac{3(y_{1}-y_{2})(z_{1}+z_{2})}{r^{^{*}5}}\bigg]$\tabularnewline
		\hline
		Top\\ plane & $\mu_{ii}^{zz}=\frac{1}{6\pi\eta b}\Big(1-\frac{3b}{4(L-z_{i})}\Big)$. & $\tilde{\mu}_{ii}^{zy}=0$ & $\hat{\mu}_{ii}^{yy}=\frac{1}{8\pi\eta b^{3}}\Big(1-\frac{1}{16}\frac{b^{3}}{(L-z_{i})^{3}}\Big)$\tabularnewline
		\hline
		Top\\ plane & $\mu_{12}^{xz}=\frac{1}{8\pi\eta}\bigg[\frac{(x_{1}-x_{2})(z_{1}-z_{2})}{r^{3}}-\frac{(x_{1}-x_{2})(z_{1}+z_{2}-2L)}{r^{^{*}3}}\bigg]$ & $\tilde{\mu}_{12}^{xy}=\frac{1}{8\pi\eta}\bigg[\frac{(z_{1}-z_{2})}{r^{3}}-\frac{(z_{1}+z_{2}-2L)}{r^{^{*}3}}\bigg]$ & $\hat{\mu}_{12}^{yy}=-\frac{1}{16\pi\eta}\bigg[\frac{1}{r^{3}}-\frac{3(y_{1}-y_{2})(y_{1}-y_{2})}{r^{5}}+\frac{1}{r^{*3}}-\frac{3(y_{1}-y_{2})(y_{1}-y_{2})}{r^{^{*}5}}\bigg]$\tabularnewline
		\hline
		Top\\ plane & $\mu_{12}^{zz}=\frac{1}{8\pi\eta}\bigg[\frac{1}{r}+\frac{(z_{1}-z_{2})(z_{1}-z_{2})}{r^{3}}-\frac{1}{r^{*}}-\frac{(z_{1}+z_{2})(z_{1}+z_{2}-2L)}{r^{^{*}3}}\bigg]$ & $\tilde{\mu}_{12}^{yz}=\frac{1}{8\pi\eta}\bigg[\frac{(x_{1}-x_{2})}{r^{3}}-\frac{(x_{1}-x_{2})}{r^{^{*}3}}\bigg]$ & $\hat{\mu}_{12}^{zy}=-\frac{1}{16\pi\eta}\bigg[\frac{1}{r^{3}}-\frac{3(y_{1}-y_{2})(z_{1}-z_{2})}{r^{5}}-\frac{1}{r^{*3}}+\frac{3(y_{1}-y_{2})(z_{1}+z_{2}-2L)}{r^{^{*}5}}\bigg]$\tabularnewline
		\hline
	\end{tabular}\caption{Explicit forms of mobility matrices near the bottom and top parallel
		no-shear planes, separated by a distance $L$. The first three rows
		contain the near bottom plane expressions which satisfy the no-shear boundary
		condition. Here $G_{\alpha\beta}(x_{1},z_{1};x_{2},z_{2})=G_{\alpha\beta}^{o}(x_{1}-x_{2},z_{1}-z_{2})+(\delta_{\beta\rho}\delta_{\rho\gamma}-\delta_{\beta3}\delta_{3\gamma})G_{\alpha\gamma}^{o}(x_{1}-x_{2},z_{1}+z_{2})$
		is the Green's function of Stokes equation which satisfy the no-shear
		condition at a plane surface \citep{aderogba1978action}, $\rho$
		takes values $x,\,y$, which correspond to directions in the plane
		surface and $G_{\alpha\beta}^{o}(\boldsymbol{r})=\frac{1}{8\pi\eta}\left(\nabla^{2}\delta_{\alpha\beta}-\nabla_{\alpha}\nabla_{\beta}\right)r$
		is the Oseen tensor. The vectors $r=\sqrt{(x_{1}-x_{2})^{2}+(z{}_{1}-z_{2})^{2}}$
		and $r^{*}=\sqrt{(x_{1}-x_{2})^{2}+(z{}_{1}+z_{2})^{2}}$ are, respectively,
		inter-colloidal distance and the distance from one colloid to the other's image charge. The last three rows are the near
		top plane expression. We emphasize that we do not include any periodic
		boundary condition and the particles are allowed to explore the full
		space between the planes.\label{tab:explicitMu}}
\end{table}

\section{Exact solution for Hamiltonian limit cycle\label{sec:Exact-solution-for}}

The two-body dynamics is described in Eqs.(\ref{eq:dynamicalSystem-1}-\ref{eq:dynamicalSystem}) of the main
text. In an unbounded domain these are simplified to the form
\[
\frac{d\theta}{mgx/8\pi\eta r^{3}}=\frac{dx}{2v_{A}\sin\theta}=\frac{dh}{v_{A}\cos\theta-\frac{mg}{8\pi\eta}\left(\frac{4}{3b}+\frac{1}{r}+\frac{z^{2}}{r^{3}}\right)}=dt.
\]
All the remaining variables are not dynamical. In particular, the
separation $z$ between the particles now remains constant. In this
limit, we obtain an integral of the motion
\begin{equation}
H(x,\theta)=\frac{mg}{8\pi\eta}\frac{1}{\sqrt{x^{2}+z^{2}}}+2v_{A}\cos\theta.\label{eq:H}
\end{equation}
We denote the level sets as $H(x,\theta)=E$. We now use the fact that
$z$ is a constant and perform the following substitutions
\begin{gather}
x=z\tan\phi,\qquad mg=6\pi\eta bv_{0}.\label{eq:xsub}
\end{gather}
We can then find the time integrals $\int^{t}O\,dt'$ for any quantity
$O\left(\phi(t),E\right)$ that can be expressed in the form $c_{0}+c_{1}\cos\phi+c_{2}\cos^{2}\phi+c_{3}\cos^{3}\phi$.
Throughout, we use the variable substitution $\int^{T_{E}}O\,dt=\int^{x_{m}}\frac{O}{\dot{x}}\,dx=\int^{\phi_{x}}\frac{O}{\dot{\phi}}\,d\phi$,
where $x_{m}=z\tan\phi_{m}$ and $x_{m}$ is the maximum amplitude
such that $E=\frac{mg}{8\pi\eta}\frac{1}{\sqrt{x_{m}^{2}+z^{2}}}+2v_{A}$.
Under these substitutions the integrals of interest take the form
\begin{alignat}{1}
\int^{\phi}\frac{c_{0}+c_{1}\cos\phi'+c_{2}\cos^{2}\phi'+c_{3}\cos^{3}\phi'}{\cos^{2}\phi'\sqrt{\left(a_{1}-\cos\phi'\right)\left(\cos\phi'-a_{2}\right)}}d\phi.\label{eq:integral}
\end{alignat}
We can then find an exact solution as a linear combination $\alpha\mathcal{F}+\beta\mathcal{E}+\gamma\text{\ensuremath{\Pi}}_{1}+\delta\text{\ensuremath{\Pi}}_{2}+\epsilon\mathcal{G}$
of elliptic integrals \citep{Grad&Rhy} and a 5\textsuperscript{th} basis function G
(see Table \eqref{tab:The-5-basis}). These integrals then become
\begin{alignat}{1}
\int\frac{{\displaystyle \sum_{i=0}^{3}}c_{i}\left(1-\left(n\sin u\right)^{2}\right)^{i}\left(1+\left(n\sin u\right)^{2}\right)^{3-i}}{\left(1-\left(n\sin u\right)^{2}\right)^{2}\left(1+\left(n\sin u\right)^{2}\right)\sqrt{\left(1+\left(\frac{n}{m}\sin u\right)^{2}\right)}}du & ,\label{eq:transformed-eqn}
\end{alignat}
where we have used the definitions
\begin{figure}
	\includegraphics[width=0.45\columnwidth]{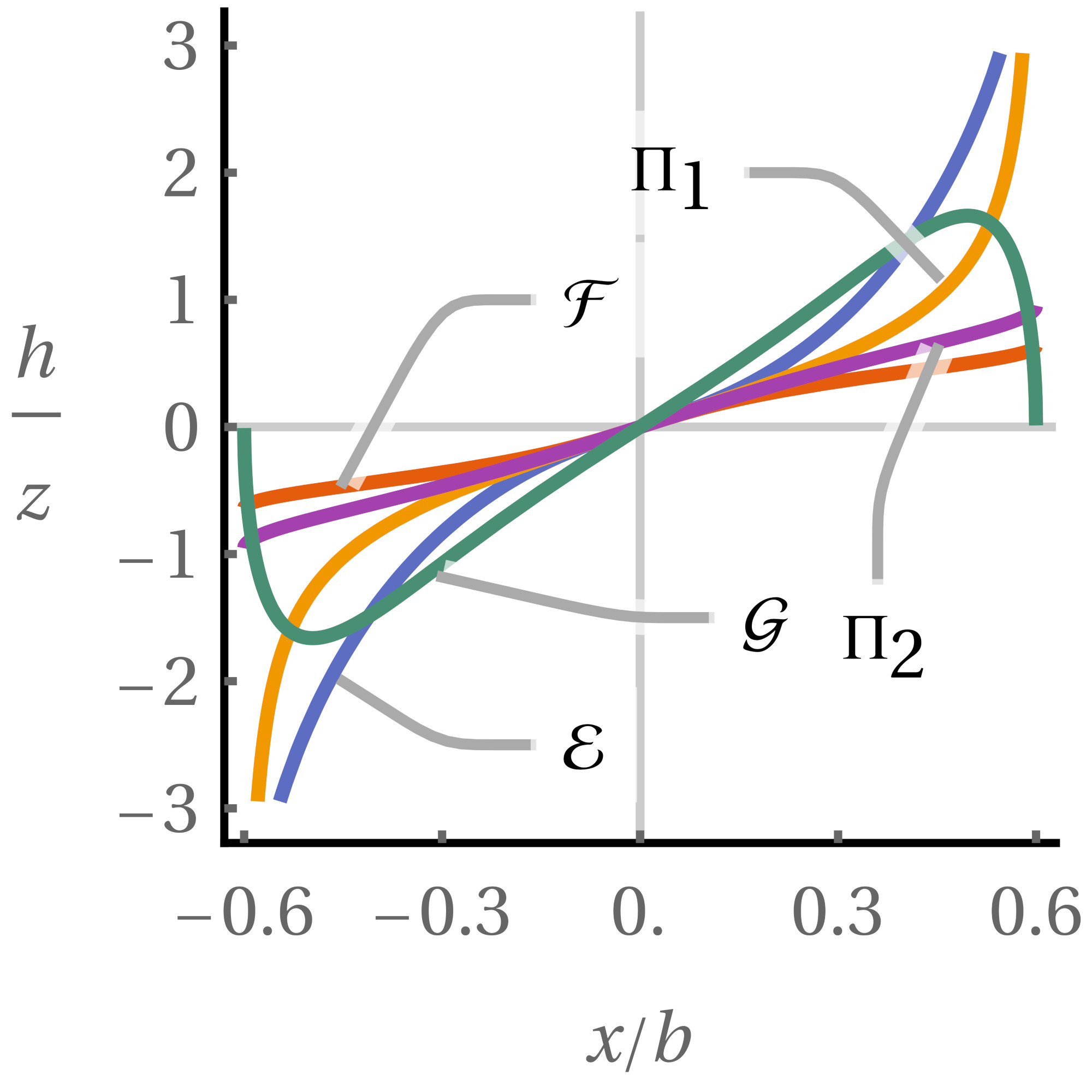}
	
	\caption{\label{fig:basisFunction}Plots of the 5 basis functions defined in
		Table (\ref{tab:The-5-basis}) for $E=2.245$.}
\end{figure}
\begin{alignat*}{1}
\tan\frac{\phi}{2} & =n\sin u,\qquad n=\sqrt{\frac{1-a_{2}}{1+a_{2}}},\qquad m=\sqrt{\frac{a_{1}-1}{a_{1}+1}}.
\end{alignat*}
By comparison with Table \eqref{tab:The-5-basis} it is easy to see that
a linear combination of the 5 functions will span the space of the
integrand in Eq.\eqref{eq:transformed-eqn}. The coefficients $(\alpha\dots\epsilon)$
are given by
\begin{gather*}
\begin{aligned}\alpha= & -\frac{c_{0}}{a_{2}}-c_{1}+c_{2}-c_{3},\quad\beta=\frac{(a_{1}-1)(a_{2}+1)c_{0}}{2a_{1}a_{2}},\quad\gamma=c_{0}\left(\frac{1}{a_{1}}+\frac{1}{a_{2}}\right)+2c_{1},\quad\delta=2c_{3},\quad\epsilon=-\frac{(a_{1}-1)(a_{2}-1)c_{0}}{4a_{1}a_{2}}\end{aligned}
\end{gather*}

To find the height $h(x,E)$ we start from the dynamical systems and
transform $x$ using Eq.(\ref{eq:xsub}) to get
\begin{align*}
h\left(x,E\right) & =\int^{\phi}\frac{\frac{E}{2}-v_{0}-\frac{9v_{0}}{8z}\cos\phi'-\frac{3v_{0}}{4z}\cos^{3}\phi'}{\cos^{2}\phi'\sqrt{4v_{A}^{2}-\left(E-\frac{3v_{0}}{4z}\cos\phi'\right)^{2}}}\,d\phi'\\
& =\int^{\phi}\frac{\frac{E}{2}-v_{0}-\frac{9v_{0}}{8z}\cos\phi'-\frac{3v_{0}}{4z}\cos^{3}\phi'}{\cos^{2}\phi'\sqrt{\left(2v_{A}-E+\frac{3v_{0}}{4z}\cos\phi'\right)\left(2v_{A}+E-\frac{3v_{0}}{4z}\cos\phi'\right)}}\,d\phi'.
\end{align*}
The above integral is of the form given in Eq.(\ref{eq:integral}),
and thus, can be rendered in the analytic form
\[
\frac{4z}{3v_{0}}\left(\alpha\mathcal{F}+\beta\mathcal{E}+\gamma\text{\ensuremath{\Pi}}_{1}+\delta\text{\ensuremath{\Pi}}_{2}+\epsilon\mathcal{G}\right),
\]
\begin{gather*}
c_{0}=\left(\frac{E}{2}-v_{0}\right),\quad c_{1}=-\frac{9v_{0}}{8z},\quad c_{2}=0,\quad c_{3}=-\frac{3v_{0}}{4z},\quad a_1=\frac{4z}{3v_{0}}\left(2v_{A}+E\right),\quad a_2=\frac{4z}{3v_{0}}\left(2v_{A}-E\right)
\end{gather*}
The constant of integration can be set to 0 w.l.o.g due to translational
invariance in the $z$ direction in the unbounded domain. We can also
calculate other useful quantities such as the average sedimentation
velocity, the time period of the oscillation and the maximum $h$
amplitude of the closed orbits seen in a co-sedimenting frame of reference
\begin{align*}
\left\langle \dot{h}\right\rangle  & =\frac{1}{T_{E}}4\int_{0}^{\phi_{\text{m}}}\frac{\dot{\tilde{h}}}{\dot{\phi}}d\phi=4h\left(\phi_{m},E\right)\\
T_{E} & =4\int_{0}^{\phi_{m}}\frac{d\phi}{\dot{\phi}}=4\int_{0}^{\phi_{\text{m}}}\frac{d\phi}{\sqrt{\left(2v_{A}-E+\frac{3v_{0}}{4z}\cos\phi\right)\left(2v_{A}+E-\frac{3v_{0}}{4z}\cos\phi\right)}\cos^{2}\phi}\\
\Delta h & =h\left(\phi_{0},E\right)-\int_{0}^{\phi_{\text{0}}}\left\langle \dot{h}\right\rangle \frac{d\phi}{\dot{\phi}}
\end{align*}
where $\phi_{0}$ is the solution of $\frac{E}{2}-v_{0}-\frac{9v_{0}}{8z}\cos\phi_{0}-\frac{3v_{0}}{4z}\cos^{3}\phi_{0}=0$.
In each case the coefficients for the $c_{i}$s can be written down
and hence the integral evaluated using Table \eqref{tab:The-5-basis}.
\begin{table}[H]
	\renewcommand{\arraystretch}{2}\centering
	
	\begin{tabular}{|>{\centering}p{0.5cm}|>{\centering}p{2.5cm}|>{\centering}p{6cm}|>{\centering}p{8cm}|}
		\hline
		& Function & Derivative & Common denominator\tabularnewline
		\hline
		\hline
		$\text{\ensuremath{\mathcal{F}}}$ & $\mathcal{F}\left(u,-\frac{n^{2}}{m^{2}}\right)$ & $\frac{1}{\sqrt{1+\frac{n^{2}}{m^{2}}\sin^{2}(u)}}$ & $\frac{\left(1-n^{2}\sin^{2}u\right)^{2}\left(1+n^{2}\sin^{2}u\right)}{\left(1-n^{2}\sin^{2}u\right)^{2}\left(1+n^{2}\sin^{2}u\right)\sqrt{1+\frac{n^{2}}{m^{2}}\sin^{2}(u)}}$\tabularnewline
		\hline
		$\mathcal{E}$ & $\mathcal{E}\left(u,-\frac{n^{2}}{m^{2}}\right)$ & $\sqrt{1+\frac{n^{2}}{m^{2}}\sin^{2}(u)}$ & $\frac{\left(1-n^{2}\sin^{2}u\right)^{2}\left(1+n^{2}\sin^{2}u\right)\left(1+\frac{n^{2}}{m^{2}}\sin^{2}u\right)}{\left(1-n^{2}\sin^{2}u\right)^{2}\left(1+n^{2}\sin^{2}u\right)\sqrt{1+\frac{n^{2}}{m^{2}}\sin^{2}u}}$\tabularnewline
		\hline
		$\text{\ensuremath{\Pi}}_{1}$ & $\text{\ensuremath{\Pi}}\left(n^{2};u,-\frac{n^{2}}{m^{2}}\right)$ & $\frac{1}{\left(1-n^{2}\sin^{2}(u)\right)\sqrt{1+\frac{n^{2}}{m^{2}}\sin^{2}(u)}}$ & $\frac{\left(1-n^{2}\sin^{2}u\right)\left(1+n^{2}\sin^{2}u\right)}{\left(1-n^{2}\sin^{2}u\right)^{2}\left(1+n^{2}\sin^{2}u\right)\sqrt{1+\frac{n^{2}}{m^{2}}\sin^{2}u}}$\tabularnewline
		\hline
		$\text{\ensuremath{\Pi}}_{2}$ & $\text{\ensuremath{\Pi}}\left(-n^{2};u,-\frac{n^{2}}{m^{2}}\right)$ & $\frac{1}{\left(1+n^{2}\sin^{2}(u)\right)\sqrt{1+\frac{n^{2}}{m^{2}}\sin^{2}(u)}}$ & $\frac{\left(1-n^{2}\sin^{2}u\right)^{2}}{\left(1-n^{2}\sin^{2}u\right)^{2}\left(1+n^{2}\sin^{2}u\right)\sqrt{1+\frac{n^{2}}{m^{2}}\sin^{2}u}}$\tabularnewline
		\hline
		$\text{\ensuremath{\mathcal{G}}}$ & $\frac{\sin2u\sqrt{1+\frac{n^{2}}{m^{2}}\sin^{2}u}}{1-n^{2}\sin^{2}u}$ & $\frac{2-2\left(2-n^{2}+2\frac{n^{2}}{m^{2}}\right)\sin^{2}u+6\frac{n^{2}}{m^{2}}\sin^{4}u-2n{}^{2}\sin^{6}u}{\left(1-n^{2}\sin^{2}u\right)^{2}\sqrt{1+\frac{n^{2}}{m^{2}}\sin^{2}u}}$ & $\frac{\left(1+n^{2}\sin^{2}u\right)\left(2-2\left(2-n^{2}+2\frac{n^{2}}{m^{2}}\right)\sin^{2}u+6\frac{n^{2}}{m^{2}}\sin^{4}u-2n{}^{2}\sin^{6}u\right)}{\left(1-n^{2}\sin^{2}u\right)^{2}\left(1+n^{2}\sin^{2}u\right)\sqrt{1+\frac{n^{2}}{m^{2}}\sin^{2}u}}$\tabularnewline
		\hline
	\end{tabular}\caption{\label{tab:The-5-basis}The 5 basis functions that make up integral
		in Eq.\eqref{eq:integral}. $\text{\ensuremath{\mathcal{F}},\ensuremath{\mathcal{E}},\ensuremath{\Pi}}$
		are incomplete elliptic integrals of the first, second and third kind
		respectively. These have been plotted in Fig.\eqref{fig:basisFunction}.}
\end{table}

\section{Krylov-Bogolyubov averaging of limit cycle at bottom plane}

The constant of the motion
\[
H(x,\theta)=\frac{mg}{8\pi\eta}\frac{1}{\sqrt{x^{2}+z^{2}}}+2v_{A}\cos\theta
\]
remains a constant if, to first order, perturbations introduced into
the equations of motion cancel. Then the perturbations have no effect
on the average orbital quantities. Our equations of motion are
\begin{alignat*}{1}
\dot{\theta} & =-\frac{mgx}{8\pi\eta r^{3}}-\omega_{R}\sin\theta\\
\dot{x} & =\,\,\,\,2v_{A}\sin\theta+\frac{9b^{2}v_{0}^{2}x}{32h^{2}},
\end{alignat*}
where the first term on the right hand side is the transient and the
second term is the perturbation. The average change of the $H$ over
a cycle is given by
\begin{align}
\left\langle \dot{H}\right\rangle  & =\int^{T_{E}}\frac{dH}{dt}dt=\int^{T_{E}}\nabla H\cdot\dot{\boldsymbol{x}}dt\\
& =\int^{T_{E}}\nabla H\cdot\Delta dt=\oint\frac{\nabla H\cdot\Delta}{\dot{\boldsymbol{x}}}d\boldsymbol{x}
\end{align}
where the transient parts of $\nabla H\cdot\dot{\boldsymbol{x}}$,
which are the equations of motion far from the planes, vanish by the
symplectic structure. The remaining part needs to be evaluated for
the perturbation vector
\[
\Delta=\begin{pmatrix}-\omega_{R}\sin\theta\\
\frac{9b^{2}v_{0}^{2}x}{32h^{2}}
\end{pmatrix}.
\]
We require the period average of $\nabla H\cdot\Delta/\dot{\boldsymbol{x}}$
to vanish to ensure that the average ``energy'' $E^{*}$ over a
cycle remains constant. We define $h^{*}$ to be the period averaged
height from the bottom plane. This immediately gives the condition
\begin{align*}
\left\langle \dot{H}\right\rangle  & =\langle2\omega_{R}v_{A}\sin^{2}\theta-\frac{9b^{2}v_{0}^{2}x^{2}}{32\left(x^{2}+z^{2}\right)^{3/2}h^{*2}}\rangle\\
& =\frac{4}{T_{E}}\int_{0}^{x_{m}}\frac{\omega_{R}\left[4v_{A}^{2}-\left(E^{*}-\frac{3bv_{0}}{4\sqrt{x^{2}+z^{2}}}\right)^{2}\right]-\frac{9b^{2}v_{0}^{2}x^{2}}{32\left(x^{2}+z^{2}\right)^{3/2}h^{*2}}}{\sqrt{4v_{A}^{2}-\left(E^{*}-\frac{3bv_{0}}{4\sqrt{x^{2}+z^{2}}}\right)^{2}}}dx\\
& =0
\end{align*}
which, under the transformation $x=z\tan\phi$, gives an integral
of the form given in Section \ref{sec:Exact-solution-for}. This condition
relates the ``equilibrium'' height $h^{*}$ and ``energy'' $E^{*}$.
A second condition comes from balancing levitation against sedimentation
over a cycle immediately giving
\begin{align*}
\left\langle \dot{h}\right\rangle  & =-\left(v_{0}+E\right)+\langle3v_{A}\cos\theta-v_{0}\left(\frac{3bz^{2}}{4r^{3}}-\frac{3b}{2h^{*}}\right)\rangle.\\
& =-\left(v_{0}+E\right)+\frac{4}{T_{E}}\int_{0}^{x_{m}}\frac{\frac{3}{2}\left(E^{*}-\frac{3bv_{0}}{4\sqrt{x^{2}+z^{2}}}\right)-v_{0}\left(\frac{3bz^{2}}{4r^{3}}-\frac{3b}{2h^{*}}\right)}{\sqrt{4v_{A}^{2}-\left(E^{*}-\frac{3bv_{0}}{4\sqrt{x^{2}+z^{2}}}\right)^{2}}}dx\\
& =0.
\end{align*}
again this integral can be put in the form of Eq.(\ref{eq:integral})
and thus we arrive at a second condition relating $h^{*}$ and $E^{*}$.
These can be solved simultaneously to give a unique estimate for the
orbit parameters of the limit cycle near the bottom plane. The pair
$\left(E^{*},h^{*}\right)$ is plotted as a function of $\omega_{R}$
in Fig.(\ref{fig:perturbationpic}c) in the main text.

\section{Exchange of dancing partners}

In this section, we consider two pairs of active particles near a
plane no-slip and no-shear surface. We emphasize that the qualitative
features of bound states predicted in the main text do not depend
on the no-slip or no-shear nature of the plane surface. Here, we show
that the interaction times at the no-shear surface is much longer
compared to a no-slip surface \citep{thutupalli2018FIPS} and that
one of the scattering states involves these interacting pairs exchanging
partners.

\begin{figure}
	\includegraphics[width=1\columnwidth]{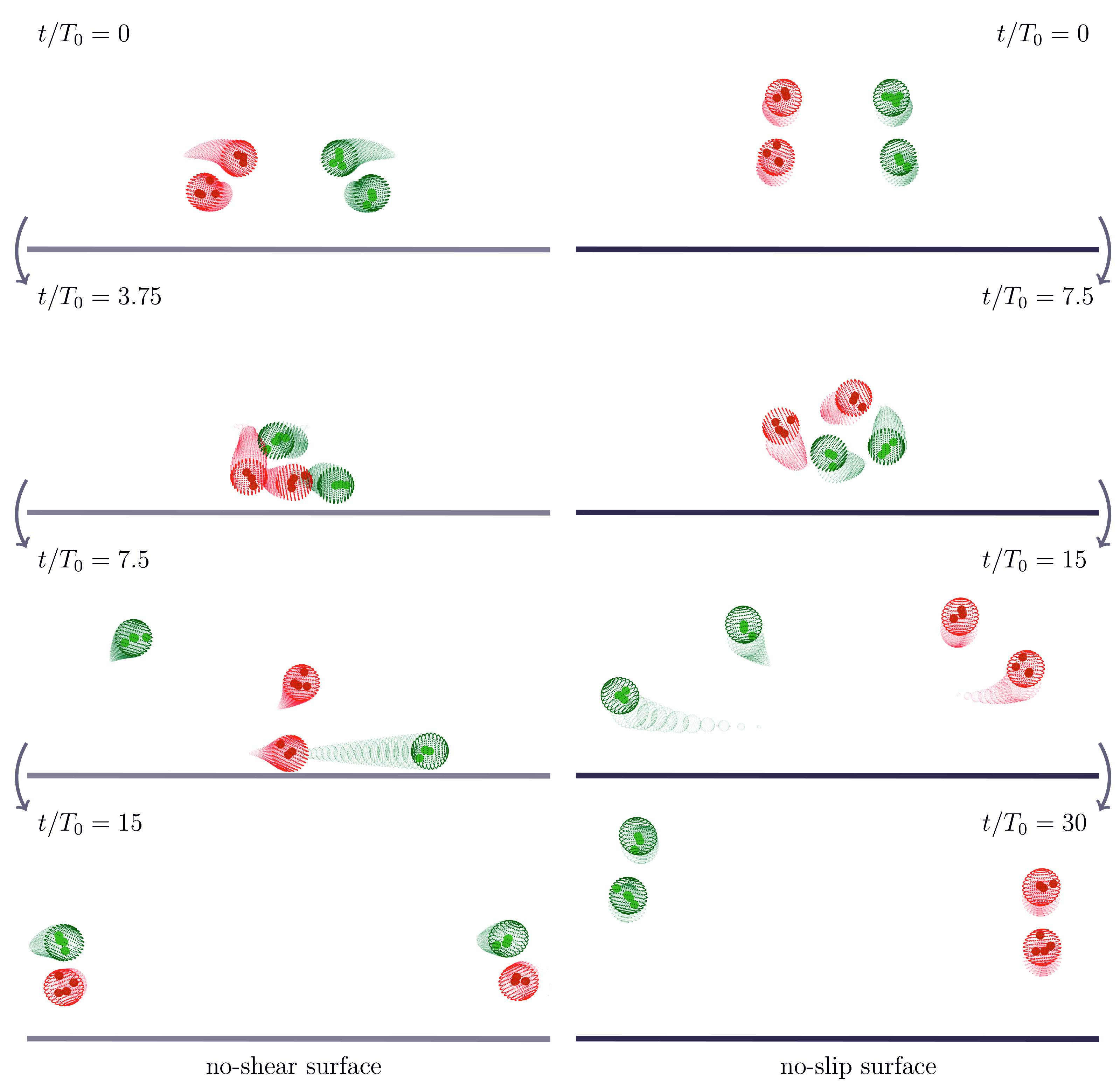}
	
	\caption{\label{fig:2Pairs}Stroboscopic images of two active bound pairs near
		a plane no-shear and no-slip surface. The presence of the no-slip
		condition at the surface weakens hydrodynamic interactions and hence
		increases the interaction time scale.\label{fig:2pair-collision}
		The particles are pulled down towards the lower surface under the
		effect of their mutual sedimentation force. Here scattering leads
		to an exchange of partners in the no-shear geometry and an exchange
		of places in the no-slip geometry. The end product of scattering events
		is highly dependent on the initial conditions. See movie 2.}
\end{figure}

The monopolar flow around an active colloid near the bottom of a parallel
plate geometry is of similar symmetry as that of a contractile dipole \citep{squiresWall},
whose axis is along the normal to the bottom surface. This has the
effect of repulsion between the bound states. These contractile flows
also produce a torque on the particles in other pairs which rotate
nearby neighbours towards one another. Active swimming is then able
to bring the two bound pairs towards one another. Repulsion dominates
when pairs are separated from each other such that $x \gg z$. On the other hand
attraction occurs if $x\sim z$. After the interaction
particles can either leave as bound pairs or single particles. Free
particles are able to swim up towards the top surface while bound
pairs stabilize near the bottom of the cell and continue their dance
indefinitely. If a no-slip surface is used instead, the individual
dancing behaviour remains the same however the inter-pair interaction
is weakened by the no-slip condition. The result is that the timescale
for pairs to come into contact is dramatically increased. Otherwise
the actual interaction and final states appear qualitatively unchanged
(see Fig.\eqref{fig:2Pairs}. We postpone further discussion of
this effect to future work. Multiparticle simulations were done with
$\omega_{R}=0.02,\,v_{0}=0.3,\,v_{A}=0.3$. The sedimentation force
was reduced in these simulations for integrator stability since the
effective hydrodynamic forces on particles becomes extremely large
when multiple particles come into close proximity.
\end{document}